\documentclass[twocolumn,
               floatfix
              ]{aastex631}

\usepackage{mathtools}
\usepackage[italicdiff]{physics}
\usepackage{savesym}
\savesymbol{tablenum}
\restoresymbol{SIX}{tablenum}
\usepackage{array}
\newcolumntype{P}[1]{>{\centering\arraybackslash}p{#1}}

\usepackage{booktabs}
\usepackage[normalem]{ulem}

\usepackage[utf8]{inputenc}
\usepackage{amsmath, amsthm, amssymb}
\usepackage{savesym}
\savesymbol{tablenum}
\restoresymbol{SIX}{tablenum}
\usepackage[italicdiff]{physics}
\usepackage{graphicx}
\usepackage[caption=false]{subfig}
\usepackage[capitalise,noabbrev]{cleveref}
\usepackage{layouts}

\shorttitle{Looking for Signs of Discreteness in the GWB}

\graphicspath{{./}{figures/}}

\begin{document}

\title{The NANOGrav 15 yr Data Set: Looking for Signs of Discreteness in the \\ Gravitational-wave Background}
  
\correspondingauthor{Andrew Casey-Clyde}
\email{andrew.casey-clyde@uconn.edu}

\author[0000-0001-5134-3925]{Gabriella Agazie}
\affiliation{Center for Gravitation, Cosmology and Astrophysics, Department of Physics, University of Wisconsin-Milwaukee,\\ P.O. Box 413, Milwaukee, WI 53201, USA}
\author[0000-0002-8935-9882]{Akash Anumarlapudi}
\affiliation{Center for Gravitation, Cosmology and Astrophysics, Department of Physics, University of Wisconsin-Milwaukee,\\ P.O. Box 413, Milwaukee, WI 53201, USA}
\author[0000-0003-0638-3340]{Anne M. Archibald}
\affiliation{Newcastle University, NE1 7RU, UK}
\author{Zaven Arzoumanian}
\affiliation{X-Ray Astrophysics Laboratory, NASA Goddard Space Flight Center, Code 662, Greenbelt, MD 20771, USA}
\author[0000-0002-4972-1525]{Jeremy George Baier}
\affiliation{Department of Physics, Oregon State University, Corvallis, OR 97331, USA}
\author[0000-0003-2745-753X]{Paul T. Baker}
\affiliation{Department of Physics and Astronomy, Widener University, One University Place, Chester, PA 19013, USA}
\author[0000-0003-0909-5563]{Bence B\'{e}csy}
\affiliation{Department of Physics, Oregon State University, Corvallis, OR 97331, USA}
\author[0000-0002-2183-1087]{Laura Blecha}
\affiliation{Physics Department, University of Florida, Gainesville, FL 32611, USA}
\author[0000-0001-6341-7178]{Adam Brazier}
\affiliation{Cornell Center for Astrophysics and Planetary Science and Department of Astronomy, Cornell University, Ithaca, NY 14853, USA}
\affiliation{Cornell Center for Advanced Computing, Cornell University, Ithaca, NY 14853, USA}
\author[0000-0003-3053-6538]{Paul R. Brook}
\affiliation{Institute for Gravitational Wave Astronomy and School of Physics and Astronomy, University of Birmingham, Edgbaston, Birmingham B15 2TT, UK}
% ...

\begin{abstract}
     \noindent The cosmic merger history of supermassive black hole binaries (SMBHBs) is expected to produce a low-frequency gravitational wave background (GWB). Here we investigate how signs of the discrete nature of this GWB can manifest in pulsar timing arrays through excursions from, and breaks in, the expected $f_{\mathrm{GW}}^{-2/3}$ power-law of the GWB strain spectrum. To do this, we create a semi-analytic SMBHB population model, fit to NANOGrav's 15 yr GWB amplitude, and with 1,000 realizations we study the populations' characteristic strain and residual spectra. Comparing our models to the NANOGrav 15 yr spectrum, we find two interesting excursions from the power-law. The first, at $2 \; \mathrm{nHz}$, is below our GWB realizations with $p$-value significance $p = 0.05$ to $0.06$ ($\approx 1.8 \sigma - 1.9 \sigma$). The second, at $16 \; \mathrm{nHz}$, is above our GWB realizations with $p = 0.04$ to $0.15$ ($\approx 1.4 \sigma - 2.1 \sigma$). We explore the properties of a loud SMBHB which could cause such an excursion. Our simulations also show that the expected number of SMBHBs decreases by three orders of magnitude, from $\sim 10^6$ to $\sim 10^3$, between $2\; \mathrm{nHz}$ and $20 \; \mathrm{nHz}$. This causes a break in the strain spectrum as the stochasticity of the background breaks down at $26^{+28}_{-19} \; \mathrm{nHz}$, consistent with predictions pre-dating GWB measurements. The diminished GWB signal from SMBHBs at frequencies above the $26$~nHz break opens a window for PTAs to detect continuous GWs from individual SMBHBs or GWs from the early universe.
\end{abstract}


\section{Introduction}
\label{sec:intro}

Massive galaxies should host supermassive black holes (SMBHs) at their centers \citep{kormendy_inward_1995}.
As galaxies merge, their central SMBHs are expected to eventually gravitationally bind, forming SMBH binaries \citep[SMBHBs, ][]{begelman_massive_1980}.
As these binaries inspiral they radiate away energy via low frequency gravitational waves (GWs). The incoherent superposition of GWs from SMBHBs forms a GW background \citep[GWB, e.g. ][]{rajagopal_ultra_1995}.

GWs from these sources, including the GWB itself, are the primary detection target of pulsar timing array (PTA) experiments \citep{foster_constructing_1990}.
Current PTA experiments, including the North American Nanohertz Observatory for Gravitational Waves \citep[NANOGrav,][]{agazie_nanograv_2023}, the European PTA (EPTA) in collaboration with the Indian PTA \citep[InPTA;][]{eptacollaboration_second_2023}, the Parkes PTA \citep[PPTA,][]{reardon_search_2023}, and the Chinese PTA \citep[CPTA,][]{xu_searching_2023}, have now reported evidence for a GWB in their pulsar data, opening a new window to study the SMBHB population.
New data from MeerKAT PTA (MPTA, \citealt{miles_meerkat_2023}) are expected to be important for GWB characterization and continuous wave (CW) searches.

Due to the slowly evolving nature of SMBHBs, we expect hundreds of thousands to millions of them to be emitting in the nanoHertz regime \citep{becsy_exploring_2022}.
The majority of these are expected to be unresolvable, however we can probe their ensemble properties, such as mass and redshift distribution, via the GWB's amplitude \citep{casey-clyde_quasarbased_2022,agazie_nanograv_2023f,eptacollaboration_second_2024}.

Moreover, there are two ways in which discrete aspects of the GWB may manifest:  firstly, a few nearby and/or very massive SMBHBs could manifest as excursions in the GWB strain spectrum before being individually resolvable with CW searches. 
Secondly, due to the discrete nature of the GWB (if indeed sourced by SMBHBs) there should be a frequency in its strain spectrum where the stochasticity of the GWB breaks down \citep{sesana_stochastic_2008,ravi_does_2012,mingarelli_characterizing_2013,roebber_cosmic_2016,kelley_gravitational_2017}. 
Here we model and search for both excursions in the strain and residual spectra, and a break in the GWB's power-law behavior at high frequencies. 

We use a state-of-the-art, semi-analytic SMBHB model to calculate the expected properties of the GWB to search for these signs of discreteness. Searching for a knee, we fit the expected GWB strain spectrum with a double power-law model. 
We then compare the expected GWB signal to the actual timing residual spectrum observed by NANOGrav \citep{agazie_nanograv_2023}, searching for excursions from the power-law behavior in the measured GWB spectra.

This paper is organized as follows:
In \autoref{sec:population} we briefly review both analytic and discrete methods for modeling the GWB.
In \autoref{sec:models} we present the analytic models that we fit to our discretely sampled GWB spectra.
In \autoref{sec:results} we present the results of these fits, including the frequency where the spectrum diverges from the expected power-law slope.
We also compare our discretely sampled GWB spectra to the spectrum observed by \citet{agazie_nanograv_2023} to search for evidence of discreteness in the data.
In \autoref{sec:discussion} we summarize our results and discuss their implications for both future GWB and CW searches.

Throughout this work we use units where $G = c = 1$.

\section{GWB Population Model}
\label{sec:population}

\begin{figure*}
    \centering
    \resizebox{\textwidth}{!}{
    \includegraphics[height=3cm]{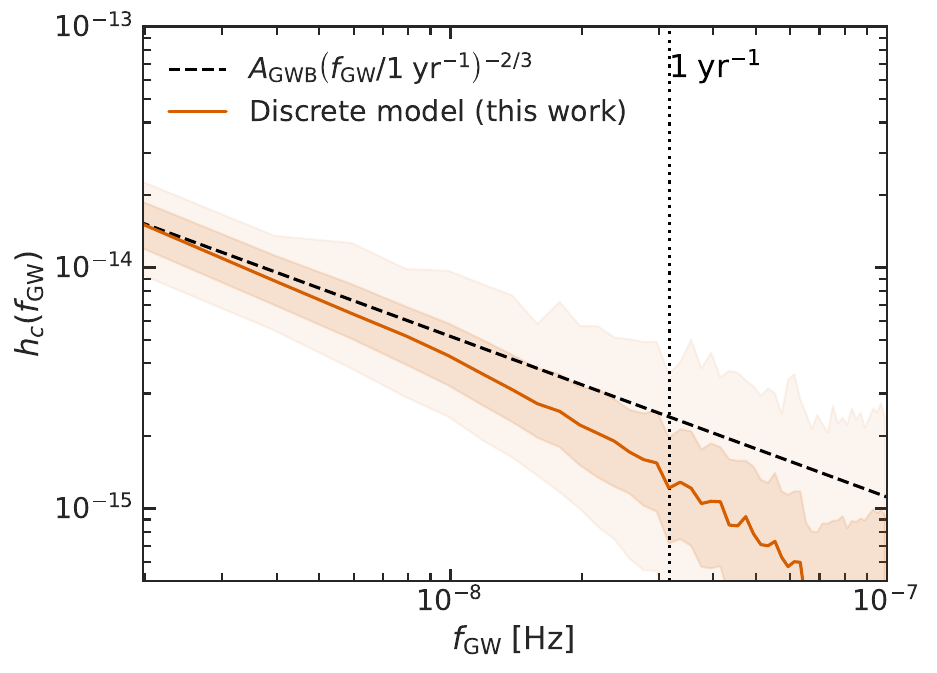}
    \includegraphics[height=2.97cm]{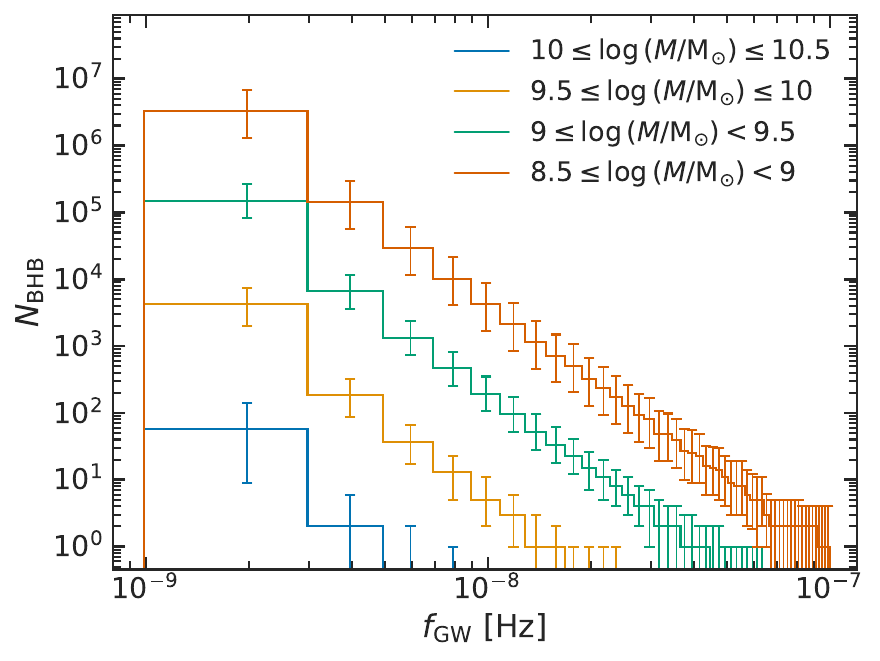}
    }
    
    \caption{
    Characteristic strain spectra of 1,000 SMBHB populations, each generating a GWB.
    \textit{Left:} The expected $h_c(f_{\mathrm{GW}})$ induced by a GWB with $A_{\mathrm{yr}} = 2.4^{+0.7}_{-0.6} \times 10^{-15}$ at $f = 1 \; \mathrm{yr}^{-1}$(red).
    The solid line shows the median GWB, while the inner and outer shaded regions denote the $68 \%$ and $95 \%$ confidence intervals, respectively.
    These confidence intervals reflect uncertainty from NANOGrav's GWB measurement and from differences between individual GWB realizations.
    At high frequencies the discretely generated GWB diverges from the power-law behavior of the analytic spectrum (black dashed line) which does not account for the discrete nature of SMBHBs.
    \textit{Right:} The number of SMBHBs generated at each frequency. Histograms show the median number of binaries generated in each frequency interval, with colors denoting mass range. Error bars denote $68 \%$ confidence intervals.
    Colored lines show the number of binaries at each frequency broken down by mass. It is clear that the vast majority of binaries at any $f_{\mathrm{GW}}$ have low masses.
    }
    \label{fig:discretized_model}
\end{figure*}

We adopt a SMBHB population model derived from the major merger rate of galaxies \citep{chen_constraining_2019}.
Briefly, in this model the galaxy major merger rate assumes a galaxy stellar mass function, an observed galaxy pairing fraction, and a theoretical galaxy merger timescale.
We then compute the SMBHB merger rate from the galaxy merger rate by assuming empirical scaling relations between galaxy stellar mass, bulge mass, and SMBH mass \citep{chen_constraining_2019}.
The SMBHB merger rate, $\dot{\phi}_{\rm BHB} = \dd[3]{\Phi_{\rm BHB}} / (\dd{M} \dd{q} \dd{t_{r}})$, describes the differential comoving number density of SMBHB mergers, $\Phi_{\rm BHB}$, per unit proper time, $t_{r}$, binary total mass, $M = M_{1} + M_{2}$, and mass ratio, $q = M_{2} / M_{1} \leq 1$, where $M_{1}$ and $M_{2}$ are the masses of the primary and secondary SMBHs in the binary, respectively.

The SMBHB merger rate is related to the characteristic strain, $h_{c}(f_{\mathrm{GW}})$, of the GWB by \citep{phinney_practical_2001}
\begin{equation}
\label{eq:gwb}
     h_{c}^{2}(f_{\mathrm{GW}}) = \frac{4}{3 \pi}\frac{1}{f_{\mathrm{GW}}^{4 / 3}} \iiint \dot{\phi}_{\rm BHB} \frac{\dd{t_{r}}}{\dd{z} }\frac{\mathcal{M}^{5 / 3}}{(1 + z)^{1 /3}} \dd{M} \dd{q} \dd{z},
\end{equation}
where $\mathcal{M}^{5 / 3} = M^{5 / 3} q / (1 + q)^{2}$ is the chirp mass, $f_{\mathrm{GW}}$ is the observed GW frequency, $z$ is the redshift, and $dt_{\rm r} / dz$ is given by cosmology \citep{hogg_distance_1999}.
\autoref{eq:gwb} is frequently written as a power-law, $h_{c}(f_{\mathrm{GW}}) = A_{\rm yr} \left(f_{\mathrm{GW}} / f_{\rm yr} \right)^{\alpha}$, where $A_{\rm yr}$ is the GWB amplitude at a reference frequency $f_{\rm yr} = 1~\rm{yr^{-1}}$.
For circular SMBHBs and GW-driven evolution, we expect $\alpha = - 2 / 3$~\citep{phinney_practical_2001}.
While \autoref{eq:gwb} can be used to model the GWB analytically, it does not account for the fact that the GWB arising from SMBHBs is composed of discrete sources.
This is important at high frequencies, where we expect the stochastic, power-law behavior of the real GWB to break down as finite number effects cause the GWB spectrum to decrease more steeply than \autoref{eq:gwb} predicts.

To model these effects we generate 1,000 realizations of discretized SMBHB populations \citep{sesana_stochastic_2008}.
We can then compute realistic GWB spectra built up from the individual SMBHB strains in each population realization.
Specifically, each population realization is generated by sampling SMBHBs from the differential number of SMBHBs per unit $M$, $q$, $z$, and $f_{\mathrm{GW}}$ \citep{chen_constraining_2019},
\begin{equation}
\label{eq:diff_num}
    \mathcal{N}_{\rm BHB} \equiv \frac{\dd[4]{N_{\mathrm{BHB}}}}{\dd{M} \dd{q} \dd{z} \dd{f_{\mathrm{GW}}}} = \dot{\phi}_{\rm BHB} \frac{dV_{\mathrm{c}}}{dz} \frac{dt_{\rm r}}{df_{\rm r}} \frac{df_{\rm r}}{df_{\mathrm{GW}}},
\end{equation}
where $dV_{\mathrm{c}} / dz$ is the comoving volume per unit redshift \citep{hogg_distance_1999}, $f_{\rm r} = f_{\mathrm{GW}} (1 + z)$ is the rest frame GW frequency, and $dt_{\rm r} / df_{\rm r} = (5 / 96) \pi^{-8 / 3} \mathcal{M}^{-5 / 3} f_{\rm r}^{-11 / 3}$ is the differential residence timescale per rest frame frequency \citep{peters_gravitational_1963}.
We use a model of $\dot{\phi}_{\rm BHB}$ derived from galaxy major merger rates which has been fit via \autoref{eq:gwb} to the GWB amplitude measured in NANOGrav's 15 yr data set, $A_{\mathrm{yr}} = 2.4^{+0.7}_{-0.6} \times 10^{-15}$, using MCMC \citep{agazie_nanograv_2023,casey-clyde_quasars_2024a}.
The resulting GWB realizations also reflect uncertainty in $A_{\mathrm{yr}}$.
The SMBH mass function implicit in our model is consistent with the local SMBH mass function (e.g., \citealp{marconi_local_2004,shankar_selfconsistent_2009,vika_millennium_2009}; cf. \citealp{sato-polito_where_2023}).
In the right side of \autoref{fig:discretized_model} we show the median number of SMBHBs generated at each frequency over all 1,000 realizations.

We compute the sky- and polarization-averaged strain, $h$, of each sampled SMBHB as \citep{thorne_gravitational_1987,sesana_stochastic_2008,rosado_expected_2015}
\begin{equation}
    h = \frac{8}{\sqrt{10}} \frac{\mathcal{M}_{\mathrm{o}}^{5 / 3} (\pi f_{\mathrm{GW}})^{2 / 3}}{D_{L}(z)},
\end{equation}
where $D_{L}(z)$ is the luminosity distance to a binary at $z$, given by standard cosmology, and where $\mathcal{M}_{\mathrm{o}} = \mathcal{M} (1 + z)$ is the observer frame chirp mass.
In each realization we then compute $h_c(f_{\mathrm{GW}})$ at each frequency via $h_{c}^{2}(f_{\mathrm{GW}}) = \sum_{k} h_{k}^{2} f_{k} / \Delta f_{\mathrm{GW}}$, where $\Delta f_{\mathrm{GW}} = 1 / T_{\rm obs}$ is the frequency sampling interval, set by the total PTA observation time, $T_{\rm obs}$.

The spectral energy density (SED) of the GWB is given by \cite{kelley_gravitational_2017}:
\begin{equation}
    S_{h}(f_{\mathrm{GW}}) = \frac{h^2_c(f_{\mathrm{GW}})}{12 \pi^2 f_{\mathrm{GW}}^3}\, .
\end{equation}
We can easily move between the characteristic strain and the residuals induced by the GWB via the amplitude spectral density (ASD), $\sqrt{S_{h}(f)/T_{\rm obs}}$.
The frequency dependence of the ASD is therefore ASD $\propto f_{\mathrm{GW}}^{-13/6}$, compared to the scaling of $h_c \propto f_{\mathrm{GW}}^{-2/3}$.
Excursions from an $f_{\mathrm{GW}}^{-2 / 3}$ power-law in characteristic strain space remain excursions from an $f_{\mathrm{GW}}^{-13/6}$ in ASD space.

\section{The High-frequency knee}
\label{sec:models}

\cite{sesana_stochastic_2008} pointed out a high-frequency GWB ``knee'' feature, where the paucity of GW sources reduces the amplitude of the strain spectrum.
Using a range of SMBHB merger rates derived from different dark matter halo merger tree prescriptions they found this knee to occur at $37^{+15}_{-13} \; \mathrm{nHz}$.
Here, we instead assume a SMBHB merger rate which has been fit to the GWB amplitude measured in NANOGrav's 15 yr data set to compute the location of the knee.
We interpret this knee frequency, $f_{\mathrm{knee}}$, as the frequency at which the stochasticity of the GWB breaks down on average, leading to deviations from an $f_{\mathrm{GW}}^{-2 / 3}$ power-law.

We consider a double power-law model of the average characteristic strain spectrum, which we implement here for the first time.
We parameterize the characteristic strain as,
\begin{equation}
\label{eq:dpl}
    h_{c}^{\rm{dpl}}(f_{\mathrm{GW}}) = \frac{2 A_{\rm knee}}{\left(\frac{f_{\mathrm{GW}}}{f_{\rm knee}}\right)^{-\alpha_{1}} + \left(\frac{f_{\mathrm{GW}}}{f_{\rm knee}}\right)^{-\alpha_{2}}},
\end{equation}
where $\alpha_{1} = -2 / 3$ is the low-frequency slope of a GWB arising from circular SMBHBs undergoing GW-driven evolution, $\alpha_{2}$ is the high frequency slope, and $A_{\rm knee}$ is the amplitude of the background at $f_{\rm knee}$.
We take $A_{\rm knee} = \lim_{f_{\mathrm{GW}} \ll f_{\mathrm{knee}}} (A_{\rm yr} / 2) (f_{\rm knee} / f_{\rm yr})^{\alpha_{1}} [1 + (f_{\mathrm{GW}} / f_{\mathrm{knee}})^{\alpha_{1} - \alpha_{2}}]$ such that \autoref{eq:dpl} reduces to the standard power-law model at $f_{\mathrm{GW}} \ll f_{\mathrm{knee}}$.

We additionally consider a physically motivated model of $h_{c}(f_{\mathrm{GW}})$ which accounts for the fact that contributions to the GWB at each $f_{\mathrm{GW}}$ must come from an integer number of SMBHBs.
Specifically, \citet{sesana_stochastic_2008} posit that in a given frequency interval there is a characteristic mass, $\tilde{M}(f_{\mathrm{GW}})$, such that less than one SMBHB with $M > \tilde{M}$ contributes to the GWB on average.
For example, in the right hand panel of \autoref{fig:discretized_model} the median number of SMBHBs with $M > 10^{10} \; \mathrm{M}_{\odot}$ in each frequency interval above $\sim 6 \; \mathrm{nHz}$ is $< 1$.
In a typical realization of the universe we therefore don't expect to have SMBHBs more massive than $10^{10} \; \mathrm{M}_{\odot}$ emitting GWs at $f_{\mathrm{GW}} \gtrsim 6 \; \mathrm{nHz}$.
By assuming that contributions to $h_{c}(f_{\mathrm{GW}})$ come from binaries with $M < \tilde{M}(f_{\mathrm{GW}})$ and that the mass dependence of $\dot{\phi}_{\mathrm{BHB}}$ can be approximated as a power law the characteristic strain of the GWB can be modeled as
\begin{equation}
\label{eq:hc_phys}
\begin{split}
    h_{c}^{\mathrm{phys}}(f_{\mathrm{GW}}) &= A_{\mathrm{yr}} \left(\frac{f_{\mathrm{GW}}}{f_{\mathrm{yr}}}\right)^{\alpha_{1}} \\
    &\qquad \times\left[1 + \left(\frac{f_{\mathrm{GW}}}{f_{\mathrm{knee}}}\right)^{11/3}\right]^{3 (\alpha_{2} - \alpha_{1}) / 11}
    \end{split}\, ,
\end{equation}
where $\alpha_{1} = -2/3$ is the slope at $f_{\mathrm{GW}} \ll f_{\mathrm{knee}}$, and where $\alpha_{2}$ is the slope of the average strain spectrum at $f_{\mathrm{GW}} \gg f_{\mathrm{knee}}$ (see Appendix~\ref{sec:physical_model} and \citealt{sesana_stochastic_2008} for details).

We constrain the double power-law and the physical models via MCMC sampling.
We use Gaussian kernel density estimators (KDEs) to estimate the probability density function (PDF) of the $h_{c}(f_{k})$ realizations at $k = 1, \ldots, 30$ frequencies, $f_{k} \approx k / 15 \; \mathrm{yr}^{-1}$. These frequencies were chosen to match the GWB spectrum frequencies from NANOGrav's 15 yr data~\citep{agazie_nanograv_2023}.
The KDE at $f_{k}$ estimates the PDF as a sum of Gaussian kernels centered on each of the $h_{c, i}(f_{k})$ realizations \citep{rosenblatt_remarks_1956,parzen_estimation_1962}:
\begin{equation}
\begin{split}
        \mathrm{PDF}\left[h_{c}(f_{k})\right] &= \frac{1}{n \sigma_{k} \sqrt{2 \pi}} \\
        & \quad \times \sum_{i = 1}^{n} \exp\left\{-\frac{1}{2} \frac{\left[h_{c}(f_{k}) - h_{c, i}(f_{k})\right]^{2}}{\sigma_{k}^{2}}\right\}\, ,
        \end{split}
    \label{eq:hc_pdf}
\end{equation}
where $n = 1,000$ is the total number of GWB realizations and $\sigma_{k}$ is the kernel bandwidth.
We choose $\sigma_{k}$ at each frequency using cross-validated grid searches, which maximize the likelihood each KDE could have generated the $h_{c, i}(f_{k})$ realizations \citep{pedregosa_scikitlearn_2011,geron_handson_2017}.
The KDEs are then used to calculate the MCMC posterior log-likelihood during fitting.

\section{Results}
\label{sec:results}

Here we present the results of our GWB spectrum analyses. We start by estimating the knee frequency of the characteristic strain spectrum in our simulated GWB spectra. We then assess how significant single-frequency excursions from a power-law in NANOGrav's 15 yr residual spectrum are compared to the spectra we compute from discrete SMBHB populations. We next compare NANOGrav's 15 yr residual spectrum to NANOGrav's 12.5 yr residual spectrum to assess whether or not excursions in the 15 yr spectrum were present in previous datasets. Finally, we consider two hypotheses to explain an observed excursion in the 15 yr spectrum around 16 nHz: we first determine whether this excursion could be due to excess white noise in the 15 yr residual spectrum. We then consider a loud binary hypothesis, wherein the 16 nHz excursion is due to a loud SMBHB that sits below the CW detection threshold.

\subsection{The Knee}

\begin{figure}
    \centering
    \includegraphics[width=\columnwidth]{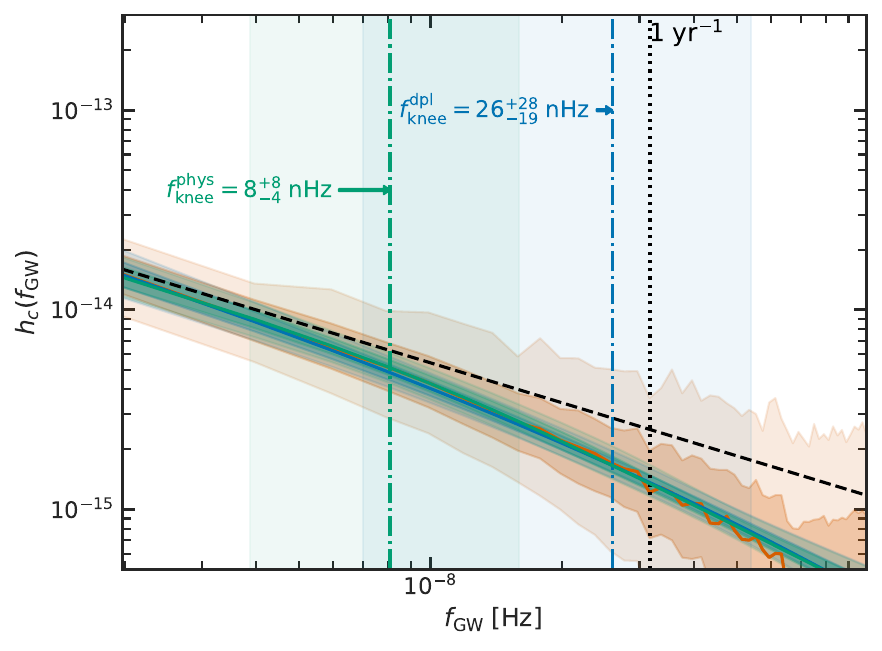}
    \caption{
    Discreteness creates a knee in the GWB strain spectrum.
    The $h_{c}(f_\mathrm{GW})$ spectrum from a population of discrete SMBHBs is shown in red, while the double power-law and physical model fits are shown in blue and green, respectively.
    The solid line shows the median value of the expected $h_{c}(f_\mathrm{GW})$, while inner and outer shaded regions show the $1 \sigma$ and $2 \sigma$ confidence intervals, as in \autoref{fig:discretized_model}.
    The dotted vertical lines show the location of the ``knee'' frequency for each fitted model, while the vertical shaded regions show the $68 \%$ confidence intervals.
    }
    \label{fig:fits}
\end{figure}

With 1,000 realizations of the cosmic population of SMBHBs, we find that the average strain spectrum clearly follows an $f_{\mathrm{GW}}^{-2/3}$ power-law at low frequencies, \autoref{fig:discretized_model}, as expected.
However, as we move to higher GW frequencies we clearly see the stochasticity of the GWB breaking down, as discrete sources start to become important.
Looking at the right side of \autoref{fig:discretized_model}, we can see that this breakdown of stochasticity is due to a dramatic decrease in the number of SMBHBs at higher frequencies.
For example, at $2 \; \mathrm{nHz}$ there are $\mathcal{O}(10^{6})$ SMBHBs contributing to the background, while at $20 \; \mathrm{nHz}$ there are fewer than $\mathcal{O}(10^{3})$ SMBHBs.
The decreasing number of SMBHBs at higher $f_{\mathrm{GW}}$ results in there being too few SMBHBs to maintain an average $f_{\mathrm{GW}}^{-2 / 3}$ power-law scaling at these higher frequencies.
This can be seen in the decreased median value of our discrete model of the GWB, as it falls below the $f_{\mathrm{GW}}^{-2 / 3}$ power-law.

In \autoref{fig:fits} we present the results of fitting the double power-law model to the generated characteristic strain spectra.
We can see that there is a bend in the strain spectrum at $f_{\mathrm{knee}} = 26^{+28}_{-19} \; \mathrm{nHz}$.
At this point the GWB deviates from the $f_{\mathrm{GW}}^{-2/3}$ power-law to $f_{\mathrm{GW}}^{-1.3^{+0.2}_{-0.3}}$, approximately twice as steep as the $f_{\mathrm{GW}}^{-2 / 3}$ power-law behavior expected at lower frequencies.

To determine if a double power-law is preferred by the data over a single power-law we calculate a Savage-Dickey Bayes factor \citep{dickey_weighted_1971}.
This can be done because the double power-law model reduces to a single power-law when $\alpha_{1} = \alpha_{2}$ (\autoref{eq:dpl}). The Bayes factor is thus computed as the ratio of the prior and posterior probability densities of $\alpha_{2}$ at $\alpha_{2} = -2/3$.
We find that a double power-law model is preferred to a single power-law with a Bayes factor of $270$.

We similarly fit a broken-power law model with $h_{c}(f_{\mathrm{GW}}) = A_{\mathrm{yr}} (f_{\mathrm{GW}} / f_{\mathrm{yr}})^{\alpha_{1}} (1 + f_{\mathrm{GW}} / f_{\mathrm{knee}})^{\alpha_{2} - \alpha_{1}}$, as in \citet{sesana_stochastic_2008}, to our 1,000 generated SMBHB spectra.
The resulting fit is nearly identical to the double power-law model, with a bend at $f_{\mathrm{knee}} = 25^{+42}_{-19} \; \mathrm{nHz}$ and a high frequency power-law slope of $f_{\mathrm{GW}}^{-1.4^{+0.3}_{-0.5}}$.
The Bayes factor for a broken power-law over a single power-law is $171$, indicating a preference for the broken power-law model.
See Appendix \ref{app:broken_pl} for details on the broken power-law model and its comparison to the double power-law model.

We also fit the physical model in \autoref{eq:hc_phys} to our generated spectra, finding $f_{\mathrm{knee}} = 8^{+8}_{-4} \; \mathrm{nHz}$ and $h_{c} \propto f_{\mathrm{GW}}^{-1.1^{+0.1}_{-0.2}}$ at $f_{\mathrm{GW}} \gg f_{\mathrm{knee}}$.
This is a lower knee than found using the double or broken power-law models, but is still within the $1 \sigma$ confidence intervals of those models.
The high frequency slope of the physical model is also shallower than the high frequency slopes of the the double and broken power-law models, but is ultimately consistent with their $1 \sigma$ confidence intervals.
The Bayes factor for the physical model over a single power-law model is $324$.
We do not attempt to characterize $f_{\mathrm{knee}}$ for NANOGrav's 15 yr GWB spectrum, which appears to be dominated by white noise above $\sim 20 \; \mathrm{nHz}$.

Finally, we compare the double power-law, broken power-law, and physical models using the Bayesian information criterion (BIC).
The BIC is defined as $\mathrm{BIC}_{X} = 2 \ln(\hat{L}_{X}) - k_{X} \ln{n}$, where $\hat{L}$ is the maximum likelihood estimate for model $X$, $k_{X}$ is the number of parameters in model $X$, and $n$ is the number of data points used for fitting \citep{schwarz_estimating_1978}.
Differences in BIC values, $\Delta(\mathrm{BIC})_{12} = \mathrm{BIC}_{1} - \mathrm{BIC}_{2}$, can be used to compare models, with positive $\Delta(BIC)_{12}$ indicating model 1 is favored over model 2 \citep{kass_bayes_1995}. 
\citet{kass_bayes_1995} suggest $\Delta(BIC)_{12} \gtrsim 10$ as a threshold for selecting model 1 over model 2.

We find the double power-law, broken power-law, and physical model all have $\mathrm{BIC} = 27$.
Thus $\Delta(\mathrm{BIC}) = 0$ between any of these models, indicating none of these models are favored over the others.
By contrast, a single power-law has $\mathrm{BIC} = 8$. 
The double power-law, broke power-law, and physical models are thus all preferred over a single power-law with $\Delta(\mathrm{BIC}) = 19$.

\subsection{Power-law Excursions}
\label{sec:pl_excursions}

\begin{figure}[t!]
    \centering
    \includegraphics[width =\columnwidth]{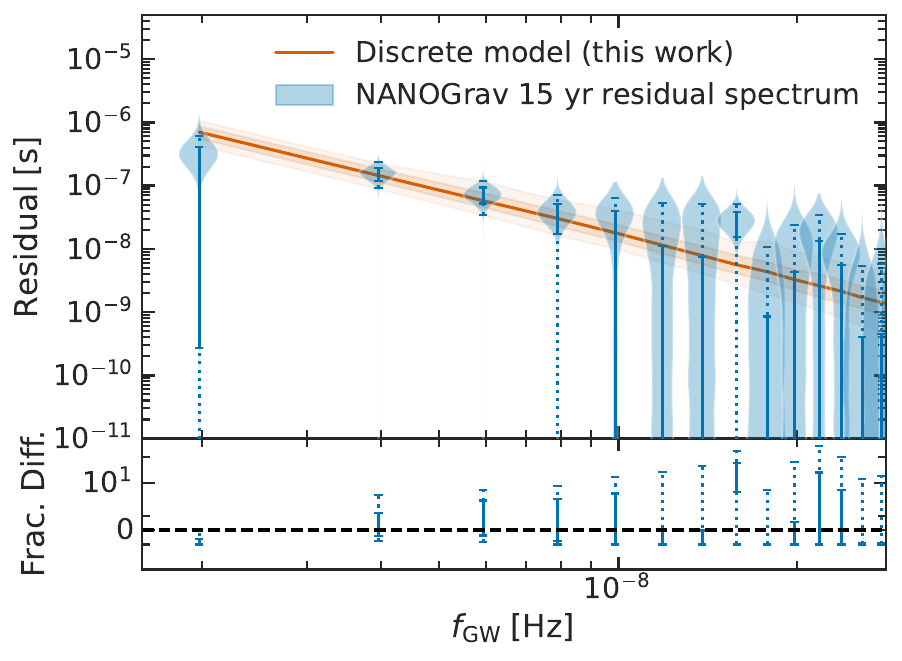}
    \caption{
    The 15 yr GWB residual spectrum compared to 1,000 spectral realizations.
    The top panel shows the observed 15 yr residual spectrum (blue) compared to 1,000 realizations of the GWB (red).
    We show the $68 \%$  and $95 \%$ confidence intervals at each frequency as solid and dotted error bars, respectively.
    We can see that these confidence intervals are heavily skewed toward small residual values.
    The bottom panel shows the fractional difference, $(S_{h}^{\mathrm{obs}} - S_{h}^{\mathrm{gen}}) / S_{h}^{\mathrm{gen}}$, between the observed SED, $S_{h}^{\mathrm{obs}}$, and that of the generated spectra, $S_{h}^{\mathrm{gen}}$.}
    \label{fig:significance}
\end{figure}

\begin{figure*}[tbh]
    \centering
     \includegraphics[width=\textwidth]{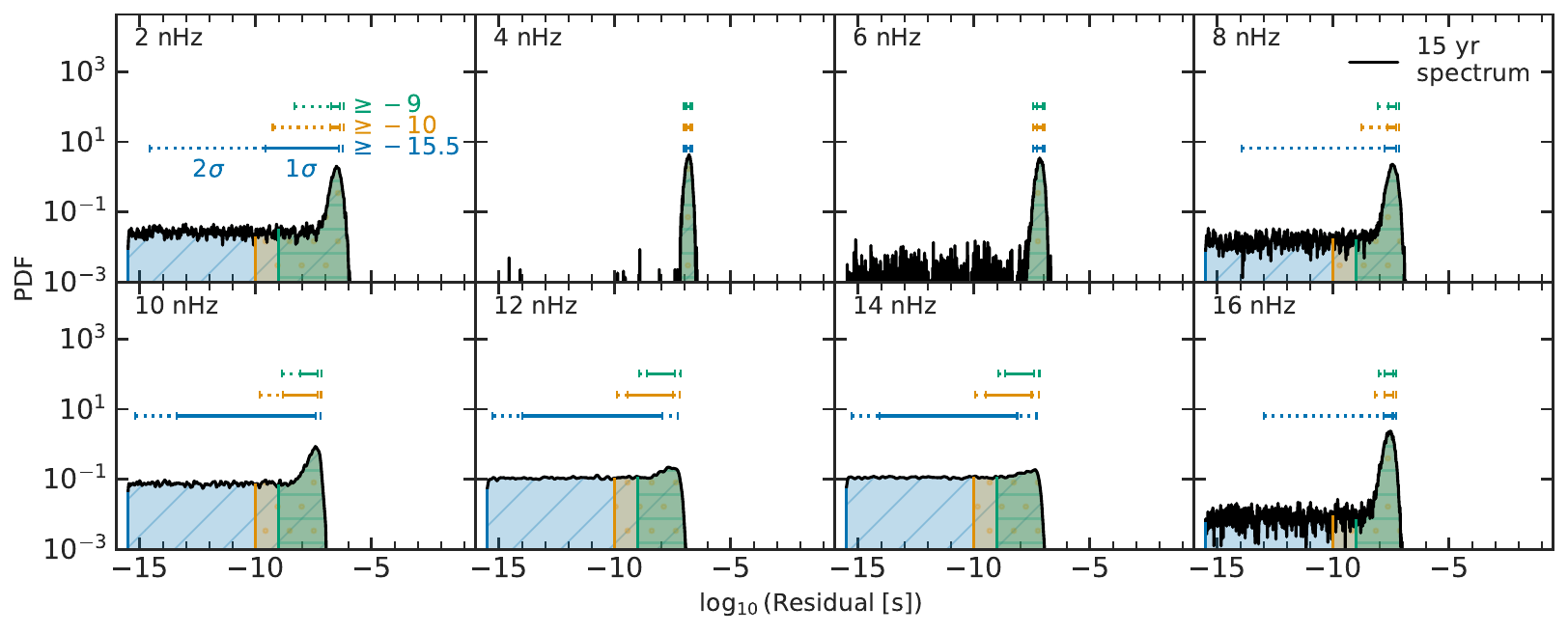}
    
    \caption{
    The broad, log-uniform support in the residual posterior distributions skews the value of the 15 yr GWB residual spectrum at each frequency. This can in turn affect our interpretation of the spectrum. Here we show a detailed view of the first eight GWB residual spectrum frequencies (black), which broadly tend towards log-uniform distributions at small residual values.
    The best constrained frequencies are $4$ and $6 \; \mathrm{nHz}$, which exhibit much lower support at small residuals than other frequencies.
    Small ``spikes" in the PDFs of these frequencies are due to the small number of MCMC samples contributing to the KDE distributions at these low residual values.
    Error bars above each distribution show the $68 \%$  and $95 \%$ confidence intervals (solid and dotted) under different posterior cuts, which are color coordinated with the shaded regions under each probability distribution.
    The original, $\log_{10} \left(\mathrm{Residual} \; [\mathrm{s}]\right) \geq -15.5$ cut is shown in blue, while a cut at $0.1 \; \mathrm{ns}$ ($\log_{10} \left(\mathrm{Residual}  \; [\mathrm{s}]\right) \geq -10$) is shown in yellow and a cut at $1 \; \mathrm{ns}$ ($\log_{10} \left(\mathrm{Residual}  \; [\mathrm{s}]\right) \geq -9$) is shown in green.
    It is clear that the choice of priors appreciably affect the long tails we see in, e.g., \autoref{fig:significance}.}
    \label{fig:free_spectrum_pdfs}
\end{figure*}

\begin{figure}
    \centering
    \includegraphics[width=\columnwidth]{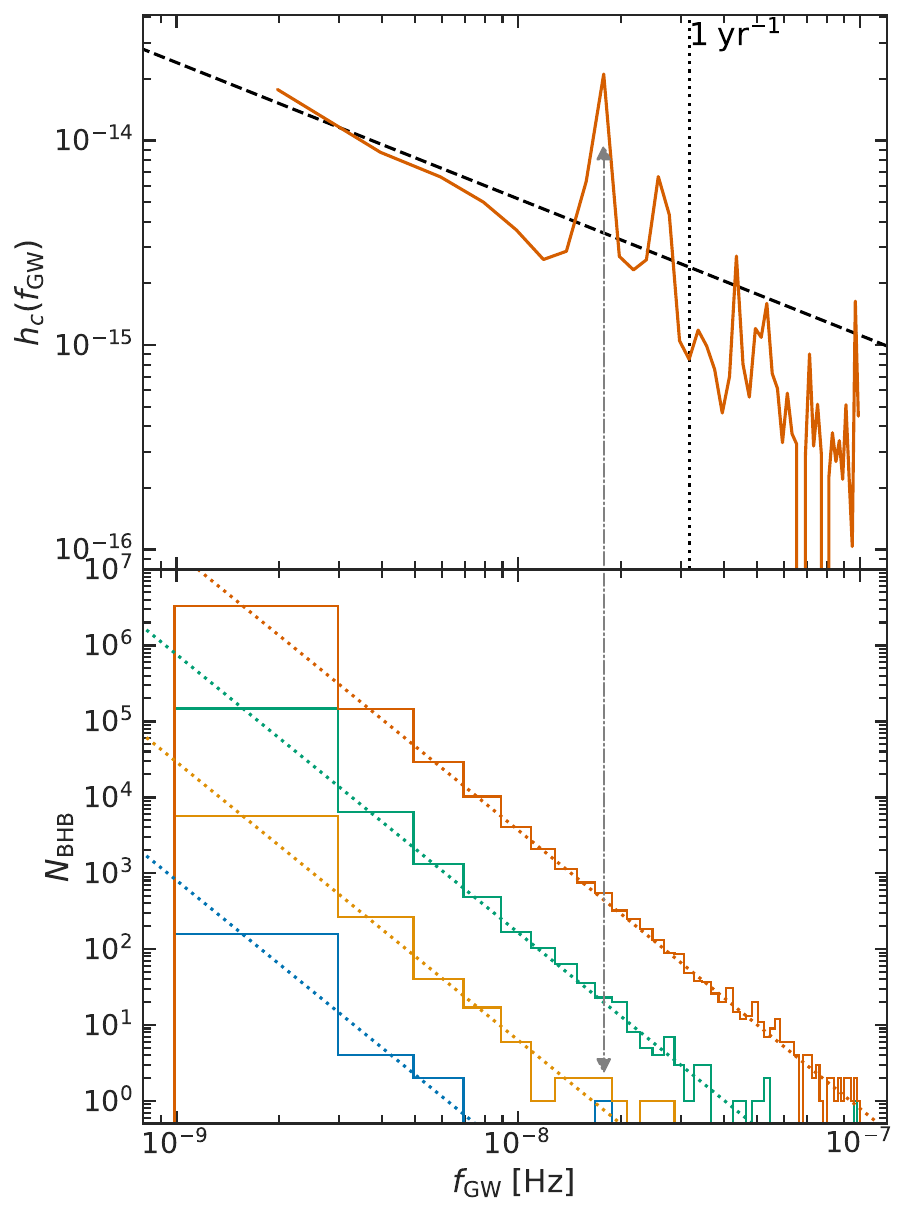}
    \caption{Example of expected excursion(s) in the GWB strain spectrum.
    \textit{Top:} A single GWB realization.
    The dashed black line shows the $f_{\mathrm{GW}}^{-2 / 3}$ power-law expectation, while the dotted gray arrow highlights one of two power-law excursions present in this realization.
    \textit{Bottom:} Corresponding histogram showing the number of SMBHBs in each frequency interval, as in the right side of \autoref{fig:discretized_model} with dotted lines showing the expected analytic distribution.
    Colors correspond to mass bins, specifically $8.5 \leq \log_{10} M < 9.0$ (red), $9.0 \leq \log_{10} M < 9.5$ (green), $9.5 \leq \log_{10} M < 10.0$ (yellow), and $10.0 \leq \log_{10} M < 10.5$ (blue).
    In this realization, the excursion below $\sim 20 \; \mathrm{nHz}$ (gray arrow) appears to come from a combination of two SMBHBs: one with $M \gtrsim 10^{9.5} \; \mathrm{M}_{\odot}$ and one with $M \gtrsim 10^{10} \; \mathrm{M}_{\odot}$. The excursion above $\sim 20 \; \mathrm{nHz}$ may result from one extra SMBHB with $M \gtrsim 10^{9.5} \; \mathrm{M}_{\odot}$.
    }
    \label{fig:example_spectrum}
\end{figure}

We next determine if excursions from $f_{\mathrm{GW}}^{-2 / 3}$ in the Hellings-Downs-correlated GWB spectrum are more signs of discreteness.
To do so we compare NANOGrav's observed 15 yr GWB spectrum to the GWB spectra we compute from discrete SMBHBs (Section~\ref{sec:population}).
We consider the difference between the observed $S_{h}$, $S_{h, k}^{\mathrm{obs}}$, and the $S_{h}$ of our generated spectra, $S_{h, k}^{\mathrm{gen}}$, at each $f_{k}$.
If the observed characteristic strain spectrum is similar to our spectral realizations, then the difference between the observed and the realizations' SEDs, $\Delta S_{h, k} = S_{h, k}^{\mathrm{obs}} - S_{h, k}^{\mathrm{gen}}$, should be zero.

The $S_{h, k}^{\mathrm{obs}}$ and $S_{h, k}^{\mathrm{gen}}$ are probability distributions -- thus the $\Delta S_{h, k}$ are also probability distributions, $P_{k}(\Delta S_{h})$.
We therefore calculate the two-sided $p$-value for the null hypothesis that $\Delta S_{h}$ is consistent with zero at each frequency, $p = 2 \min\{P_{k}(\Delta S_{h} \geq 0), P_{k}(\Delta S_{h} \leq 0)\}$.
We focus the presentation of our results on the GWB spectrum values at $2$, $12$, $14$, and $16 \; \mathrm{nHz}$, which were highlighted in \citet{agazie_nanograv_2023}, though all $30$ GWB spectrum frequencies were analyzed.
We find that the observed residual spectrum value at $2 \; \mathrm{nHz}$ is below our GWB realizations with $p = 0.05$, corresponding to a $1.9 \sigma$ excursion.
The values at $12 \; \mathrm{nHz}$ and $14 \; \mathrm{nHz}$ are below our GWB with $p = 0.31$ ($1.0 \sigma$) for each.
Finally, the spectrum value at $16 \; \mathrm{nHz}$ is above our GWB with $p = 0.15$ ($1.4 \sigma$, \autoref{fig:significance}).

We notice, however, that the $95 \%$ confidence intervals for the residual spectrum -- shown by the dotted error bars in \autoref{fig:significance} -- skew heavily towards small residuals.
We investigate this further in \autoref{fig:free_spectrum_pdfs}, which shows the log-PDFs of the residual spectrum posteriors.
We find that the residual spectrum posteriors appear to be log-uniform at $\lesssim 0.1 - 1 \; \mathrm{ns}$, reflecting the choice of prior.
This indicates that very weak GWB signals are not heavily disfavored at most frequencies the same way that very loud signals are disfavored.
Consequentially, the significance of excursions from an expected $f_{\mathrm{GW}}^{-2 / 3}$ power-law can sometimes depend on where we cut these PDFs.

To assess how the significance of excursions at $2 \; \mathrm{nHz}$, $12 \; \mathrm{nHz}$, $14 \; \mathrm{nHz}$, and $16 \; \mathrm{nHz}$ change with the prior range for the residuals, we artificially cut the residual spectrum PDFs at $0.1 \; \mathrm{ns}$ and $1 \; \mathrm{ns}$.
The lower value of $0.1 \; \mathrm{ns}$ is chosen to reflect the residual values where the PDFs appear to transition to prior dominated, log-uniform distributions.
The upper value of $1 \; \mathrm{ns}$ corresponds to the prior bound used to analyze NANOGrav's 12.5 yr dataset \citep{arzoumanian_nanograv_2020}.
To make these cuts we consider only the portion of the residual spectrum PDFs above $0.1 \; \mathrm{ns}$ and $1 \; \mathrm{ns}$. We then renormalize each PDF so they integrate to unity above these limits.

We find the residual spectrum at $2 \; \mathrm{nHz}$ remains below our GWB spectral realizations with $p = 0.05$ ($1.9 \sigma$) when using a residual cut at $0.1 \; \mathrm{ns}$, and with $p = 0.06$ ($1.8 \sigma$) when using a $1 \; \mathrm{ns}$ residual cut.
With a $0.1 \; \mathrm{ns}$ residual cut, the spectrum values at $12 \; \mathrm{nHz}$ and $14 \; \mathrm{nHz}$ are below our spectral realizations with $p = 0.78$ ($0.3 \sigma$) and $p = 0.92$ ($0.2 \sigma$), respectively. With a $1 \; \mathrm{ns}$ residual cut they are both above our spectral realizations with $p = 0.82$ ($0.2 \sigma$).
The spectrum at $16 \; \mathrm{nHz}$ lies above our GWB realizations with $p = 0.06$ ($1.9 \sigma$) using a $0.1 \; \mathrm{ns}$ residual cut, and with $p = 0.04$ ($2.1 \sigma$) using a $1 \; \mathrm{ns}$ residual cut.
All other frequencies are $\lesssim 1 \sigma$ from the median of the GWB realizations, regardless of prior choice.
Results for the $2 \; \mathrm{nHz}$, $12 \; \mathrm{nHz}$, $14 \; \mathrm{nHz}$, and $16 \; \mathrm{nHz}$ frequencies are summarized in \autoref{tab:significance}.

\begin{table}[b!]
    \centering
    \begin{tabular}[b]{|c|c|c|c|}
        \hline
        $f_{\mathrm{GW}} \; [\mathrm{nHz}]$ & $p_{15.5}$ ($\sigma_{15.5}$) & $p_{10}$ ($\sigma_{10}$) & $p_{9}$ ($\sigma_{9}$) \\
        \hline \hline
        $2$ & $0.05$ ($-1.9\sigma$) & $0.05$ ($-1.9\sigma$) & $0.06$ ($-1.8\sigma$) \\
        $12$ & $0.31$ ($-1.0\sigma$) & $0.78$ ($-0.3\sigma$) & $0.92$ ($0.1\sigma$) \\
        $14$ & $0.31$ ($-1.0\sigma$) & $0.82$ ($-0.2\sigma$) & $0.82$ ($0.2\sigma$) \\
        $16$ & $0.15$ ($1.4\sigma$) & $0.05$ ($1.9\sigma$) & $0.04$ ($2.1\sigma$) \\
        \hline
    \end{tabular}
    \caption{Significance of excursions in the observed GWB residual spectrum from the range of spectra computed from discrete SMBHB populations for the four frequencies highlighted in \citet{agazie_nanograv_2023}. $p$-values reflect the consistency of $\Delta S_{h}$ with zero, while corresponding $\sigma$ significances are given in parentheses. Subscripts $X$ on $p_{X}$ and $\sigma_{X}$ in the last three columns denote the lower log-residual cut, i.e., $\log_{10} \left(\mathrm{Residual}\right) \geq -X$. The 15 yr GWB residual spectrum at $2 \; \mathrm{nHz}$ is $\sim 2 \sigma$ below the median value of our realizations regardless of the prior cut used, while at $16 \; \mathrm{nHz}$ the spectrum is up to $2.1 \sigma$ above the median value, depending on the prior cut, \autoref{fig:free_spectrum_pdfs}. At $12 \; \mathrm{nHz}$ and $14 \; \mathrm{nHz}$ the spectrum is consistent with our GWB realizations.}
    \label{tab:significance}
\end{table}

It is worth emphasizing that analytic models of the GWB strain spectrum, including an $f_\mathrm{GW}^{-2 / 3}$ power-law and more complicated power-laws, are only statistical descriptions of the strain spectrum arising from the expected statistical distribution of SMBHBs over $f_{\mathrm{GW}}$ \citep{phinney_practical_2001,jaffe_gravitational_2003,sesana_stochastic_2008,lamb_spectral_2024}.
The spectrum of a single realization of the GWB depends only on the underlying SMBHB population.
For example, in \autoref{fig:example_spectrum} we show a single characteristic strain spectrum realization out of the 1,000 we generated.
This realization includes two excursions from the expected $f_{\mathrm{GW}}^{-2 / 3}$ power-law behavior near $20 \; \mathrm{nHz}$.
Each excursion appears to be associated with having $\mathcal{O}(1)$ more massive or nearby SMBHB than expected from na\"{i}vely extrapolating the $f_{\mathrm{GW}}^{-2 / 3}$ power-law to higher frequencies.
This demonstrates that excursions from the expected $f_{\mathrm{GW}}^{-2 / 3}$ power-law behavior of the GWB are unsurprising.
Even a single massive or nearby binary in excess of the power-law expectation can lead to excursions from an $f_{\mathrm{GW}}^{-2 / 3}$ power-law.

\vspace{1cm}
\subsection{Comparison to 12.5 yr Spectrum}

\begin{figure}
    \centering
\includegraphics[width=\columnwidth]{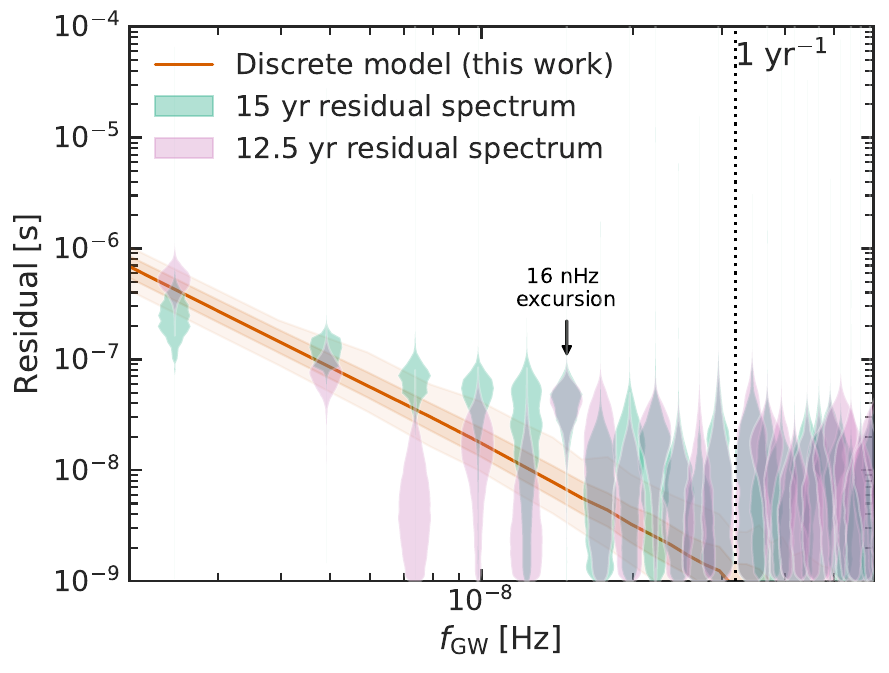}
    \caption{Comparison of 12.5 yr and 15 yr residual spectra, both using 12.5 yr frequency intervals. Specifically, we have re-computed the 15 yr residual spectrum using the 15 yr pulsar timing data with $\Delta f = 2.5 \; \mathrm{nHz}$. We can see that the excursion at $16 \; \mathrm{nHz}$ is remarkably consistent from the 12.5 yr data set to the 15 yr data set.}
    \label{fig:12p5yr_comparison}
\end{figure}

It is also interesting to compare the 15 yr residual spectrum to the residual spectrum from NANOGrav's 12.5 yr data set \citep{arzoumanian_nanograv_2020}.
Assuming the GWB is a stationary signal (i.e., it does not change appreciably from the 12.5 yr data set to the 15 yr data set), any differences between the two sets of spectra should be due to differences between the 12.5 yr and 15 yr data sets, including increased pulsar timing baselines, the inclusion of more pulsars, and improved pulsar noise modeling in the the 15 yr data set compared to the 12.5 yr data set \citep{agazie_nanograv_2023b,agazie_nanograv_2023a}.
By comparing the residual spectra we can thus determine if excursions from an expected $f_{\mathrm{GW}}^{-2 / 3}$ power-law in the 15 yr dataset are consistent with previous NANOGrav data sets or if they have only emerged recently as NANOGrav data sets have improved.

The frequency intervals in both the 12.5 yr and 15 yr residual spectra are $\Delta f_{\mathrm{GW}} = 1 / T_{\mathrm{obs}}$, corresponding to the minimum frequency resolution in each data set.
Since the 15 yr data set has a longer observation time than the 12.5 yr data set, the interval, $\Delta f_{15} = 2.0 \; \mathrm{nHz}$, in the 15 yr data set is smaller than the interval in the 12.5 yr data set, $\Delta f_{12.5} = 2.5 \; \mathrm{nHz}$.
To consistently compare the residual spectra measured in each data set we re-compute the 15 yr residual spectrum using the coarser $\Delta f_{12.5}$ intervals, \autoref{fig:12p5yr_comparison}. This is equivalent to decomposing the GWB signal in the 15 yr data set on the 12.5 yr Fourier basis frequencies, $f_{k} \approx k / 12.5 \; \mathrm{yr}^{-1}$ for $k = 1, \ldots, 30$.

We see that the $16 \; \mathrm{nHz}$ excursion remains consistently $\sim 2 \sigma$ above the median value of our GWB realizations.
At frequencies higher than $16 \; \mathrm{nHz}$ the 12.5 yr and 15 yr spectra appear to have similar white noise.
Below $16 \; \mathrm{nHz}$ the spectra differ appreciably.
These differences may be due to differences between the data sets themselves, such as $3 \; \mathrm{yr}$ longer timing baselines in the 15 yr data set compared to the 12.5 yr data set, and the addition of 21 new pulsars with $\geq 2.5 \; \mathrm{yr}$ timing baselines in the 15 yr data set.

\vspace{1cm}
\subsection{Excess White Noise Hypothesis}

We next assess the significance of excursions at $2$, $12$, $14$, and $16 \; \mathrm{nHz}$ under the assumption that NANOGrav's 15 yr residual spectrum includes excess white noise.
We compute a phenomenological residual spectrum model by adding constant white noise, $\sigma_{\mathrm{WN}}$, which is uncorrelated over $f_{\mathrm{GW}}$ to the 1,000 GWB spectra computed from discrete SMBHBs, consistent with the flat residual spectrum above $\sim 20 \; \mathrm{nHz}$.
Specifically, we model the characteristic strain spectrum of each of the 1,000 generated spectra and at each $f_{k}$ as a chi-distributed random variable with two degrees of freedom, centered on $h_{c}(f_{k})$ with scale parameter $\sigma_{\mathrm{WN}}$.
We then fit this model to NANOGrav's 15 yr residual spectrum, finding $\sigma_{\mathrm{WN}} \sim 7 \; \mathrm{ns}$.

We compare our generated residual spectra plus additional white model to NANOGrav's 15 yr residual spectrum as in Section~\ref{sec:pl_excursions}.
We find that the excursion at $2 \; \mathrm{nHz}$ is $\sim 2 \sigma$ below the discrete spectrum with additional noise ($p = 0.04 - 0.05$, depending on prior cut), the excursions at $12$ and $14 \; \mathrm{nHz}$ are $1.3\sigma$ to $0.3\sigma$ below the discrete spectrum with additional noise ($p = 0.2 - 0.8$), and the excursion at $16 \; \mathrm{nHz}$ is $\sim 1 \sigma$ above the discrete spectrum with additional noise ($p = 0.3 - 0.4$).
Full results for these frequencies are summarized in \autoref{tab:wn_significance}.

We note that NANOGrav's analysis pipeline already includes three white noise parameters encompassing different white noise sources.
The inclusion of these parameters yields a reduced $\chi^{2}$ near unity for the fit of the pulsar timing model to timing residuals without necessitating additional sources of noise \citep{agazie_nanograv_2023a}.
$7 \; \mathrm{ns}$ of excess white noise correlated between pulsars in the residual spectrum is therefore difficult to justify.
The white noise component assumed here therefore lacks a clear physical motivation.

It is possible this excess results from the fact that NANOGrav uses the maximum a posteriori white noise parameter values for each pulsar in their GWB search and characterization, as marginalizing over white noise parameters is computationally infeasible \citep{agazie_nanograv_2023a}.
Alternatively, this excess noise may be due to using a one-size-fits-all approach for pulsar noise modelling which does not consider additional sources of noise in individual pulsars \citep{lentati_spin_2016,falxa_searching_2023,larsen_nanograv_2024}.
A deeper investigation of noise in individual pulsars in NANOGrav's 12.5 yr data set is currently under way (Simon et al. in prep.).
The custom noise model techniques developed in Simon et al. (in prep.) will then be applied to the full 15 yr data set (Agazie et al. in prep.).

\begin{table}
    \centering
    \begin{tabular}[b]{|c|c|c|c|}
        \hline
        $f_{\mathrm{GW}} \; [\mathrm{nHz}]$ & $p_{15.5}$ ($\sigma_{15.5}$) & $p_{10}$ ($\sigma_{10}$) & $p_{9}$ ($\sigma_{9}$) \\
        \hline \hline
        $2$ & $0.04$ ($-2.0\sigma$) & $0.05$ ($-2.0\sigma$) & $0.05$ ($-2.0\sigma$) \\
        $12$ & $0.21$ ($-1.2\sigma$) & $0.52$ ($-0.6\sigma$) & $0.73$ ($-0.3\sigma$) \\
        $14$ & $0.21$ ($-1.3\sigma$) & $0.55$ ($-0.6\sigma$) & $0.79$ ($-0.3\sigma$) \\
        $16$ & $0.39$ ($0.9\sigma$) & $0.31$ ($1.0\sigma$) & $0.29$ ($1.1\sigma$) \\
        \hline
    \end{tabular}
    \caption{Significance of excursions in the observed GWB residual spectrum as in \autoref{tab:significance}, but with an additional $7\;\mathrm{ns}$ of excess white noise.
    }
    \label{tab:wn_significance}
\end{table}

\vspace{1cm}
\subsection{Loud Binary Hypothesis}

Another possible explanation for the excess at $16 \; \mathrm{nHz}$ is the presence of a loud SMBHB at that frequency, such as the example in \autoref{fig:example_spectrum}.
We test this hypothesis by constraining the additional strain amplitude a loud binary would contribute to the GWB to be consistent with the $16 \; \mathrm{nHz}$ excursion.
We then compare this hypothetical strain to constraints from NANOGrav's 15 yr CW search \citep{agazie_nanograv_2023d}, which found no CWs.
Other possible sources for this excursion, such as mis-modelled pulsar noise or multiple SMBHBs emitting at $16 \; \mathrm{nHz}$, will require further investigation.
See \autoref{sec:discussion} for further discussion.

\begin{figure*}
    \centering
    \includegraphics[width=1.06\columnwidth]{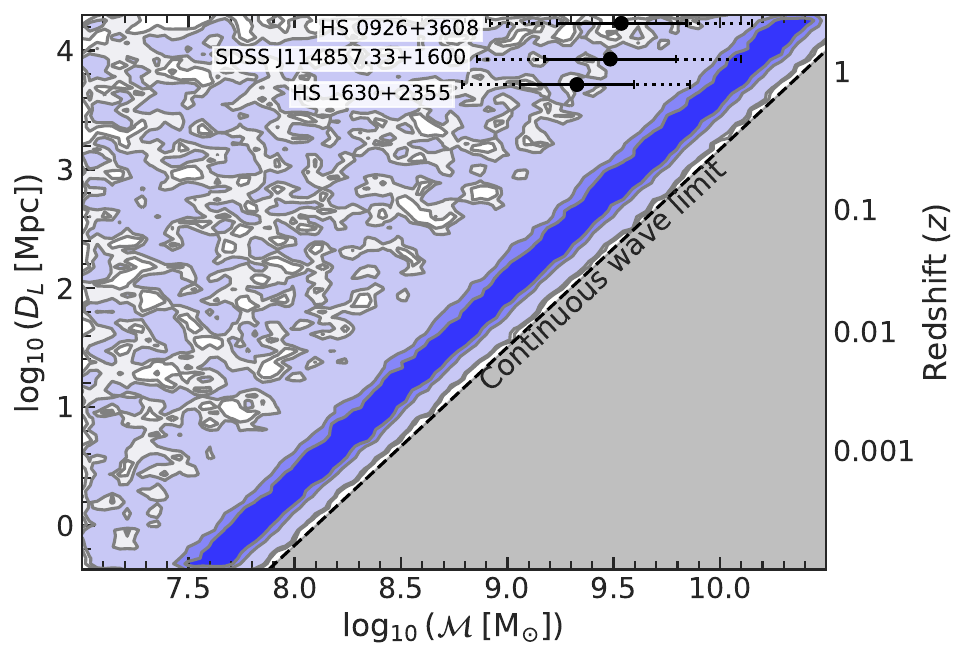}
    \includegraphics[width=.94\columnwidth]{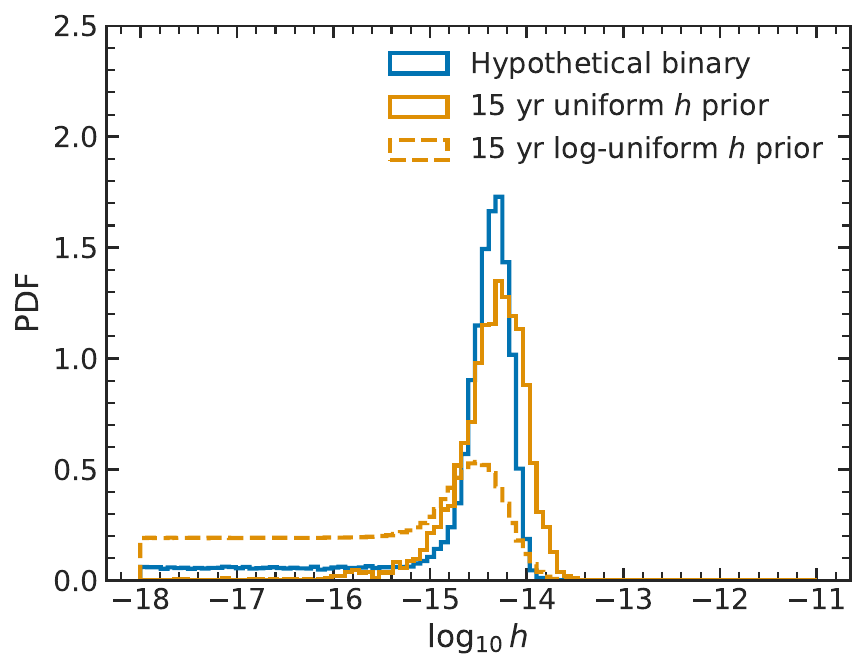}
    \caption{The 16 nHz excursion can be explained by the presence of an additional SMBHB at that frequency. \textit{Left:} Allowed luminosity distance, $D_{L}$, and chirp mass, $\mathcal{M}$, distribution of single binaries which could reproduce the excess at $16 \; \mathrm{nHz}$. Contours and blue shading show the regions of parameter space containing (from darkest to lightest) $50 \%$, $68 \%$, $95 \%$, and $99 \%$ of MCMC samples. The dashed black line and gray region show the parameter space that has already been ruled out by the non-detection of continuous waves in NANOGrav's 15 year data set \citep{agazie_nanograv_2023d}.
    Black dots show the median $\mathcal{M}$ estimates and $D_{L}$ of four SMBHB candidates. 
    Solid error bars show the $68 \%$ $\mathcal{M}$ confidence intervals for each candidate while dotted error bars show $95 \%$ $\mathcal{M}$ confidence intervals.
    \textit{Right:} The strain of a hypothetical binary at $16 \; \mathrm{nHz}$ (blue) is comparable to CW constraints from NANOGrav's 15 yr data set (yellow). NANOGrav's 15 yr CW constraints were set using two sets of priors on strain amplitude: uniform (solid yellow) and log-uniform priors (dashed yellow). Interestingly, both sets of 15 yr CW constraints have peaks near the hypothetical binary strain peak. However the differences between the hypothetical binary strain and the CW constraints are only marginal. Therefore a binary comparable to the hypothetical binary we consider here may not have been detectable in the 15 yr data set.}
    \label{fig:bin_8}
\end{figure*}

If we assume the $16 \; \mathrm{nHz}$ excursion is due to a loud binary in addition to an underlying population of fainter binaries, we find it would need to have a sky- and polarization-averaged strain of $h = \left(3.8^{+2.3}_{-3.2}\right) \times 10^{-15}$, where the central given value is the median, and upper and lower uncertainties denote the $68 \%$ confidence interval.
For comparison, the hypothetical binary CW $95 \%$ upper limit is $7.8 \times 10^{-15}$ at this frequency.
In \autoref{fig:bin_8} we show the range of $\mathcal{M}$ and luminosity distances, $D_{L}$, which could produce this strain.

Importantly, NANOGrav did not find significant evidence of CWs in the 12.5 yr and 15 yr data sets \citep{arzoumanian_nanograv_2023a,agazie_nanograv_2023d}.
Thus we compare the hypothetical binary strain to constraints on CW strain near $16 \; \mathrm{nHz}$ from NANOGrav's 15 yr CW search, \autoref{fig:bin_8}.
We consider two sets of constraints from NANOGrav's 15 yr CW search: one using uniform strain amplitude priors and one with log-uniform priors.
Interestingly, both sets of constraints have peaks near the hypothetical binary strain peak, as shown in the right side of \autoref{fig:bin_8}.
For each set of constraints we calculate the $p$-value for the null hypothesis that the 15 yr CW constraints are consistent with the hypothetical binary strain.
We find $p = 0.69$ for the uniform prior constraints and $p = 0.65$ for the log-uniform prior constraints.
Thus if the $16 \; \mathrm{nHz}$ excursion is due to a binary, it would not have been detected in NANOGrav's 15 yr CW search.

There are several extant SMBHB candidates which have $\mathcal{M}$ and $D_{L}$ that are consistent with the 16 nHz excursion.
Here we show three such candidate SMBHBs, HS 1630+2355, HS 0926+3608, and SDSS J114857.33+1600.
These were identified by \citet{xin_multimessenger_2021} as interesting candidates, due to their optical periodic quasar light curves in the Catalina Real-time Transient Survey \citep[CRTS, ][]{graham_systematic_2015}, and large masses.
Hydrodynamical simulations show that SMBHBs can induce periodicity in quasar light curves, though the specific relationship between binary periodicity and light curve periodicity is uncertain \citep{farris_binary_2014,westernacher-schneider_multiband_2021,cocchiararo_electromagnetic_2024}.
We use full-width half max binary mass estimates, $M$, from \citet{xin_multimessenger_2021} to calculate $\mathcal{M}$ with Monte Carlo uncertainty estimates, drawing the binary mass ratio $q$ from a uniform distribution between $0.1$ and $1$.
Uncertainties on $\mathcal{M}$ are dominated by the $\sim 0.3 \; \mathrm{dex}$ Gaussian uncertainties on $M$.

A SMBHB's GW emission frequency evolves as $\dot{f}_{\mathrm{GW}} = (96 / 5) \pi^{8 / 3} \mathcal{M}^{5 / 3} f_{\mathrm{r}}^{11 / 3}$, such that a binary emitting at $16 \; \mathrm{nHz}$ evolves at a rate of $\sim 5 \times 10^{-4} \left(\mathcal{M} / 10^{9} \; \mathrm{M}_{\odot}\right)^{5 / 3} \left(f_\mathrm{GW} / 16~\mathrm{nHz} \; \right)^{11/3} \mathrm{nHz} \; \mathrm{yr}^{-1}$.
At the same GW frequency, a rarer $\mathcal{M} = 10^{10} \; \mathrm{M}_{\odot}$ SMBHB (\autoref{fig:discretized_model}), will evolve at a rate of $\dot{f}_{\mathrm{GW}} \approx 0.02 \; \mathrm{nHz} \; \mathrm{yr}^{-1}$.
If a single loud binary is the source of the $16 \; \mathrm{nHz}$ excursion, we thus expect it to persist in future NANOGrav data sets.

\section{Discussion}
\label{sec:discussion}

We carried out a suite of Monte Carlo realizations to determine how the discreteness of the GWB manifests in its strain and residual spectra.
We predicted a knee in the GWB strain spectrum at $26^{+28}_{-19} \; \mathrm{nHz}$ using a double power-law model, and at $8^{+8}_{-4} \; \mathrm{nHz}$ using a physically motivated model of the characteristic strain spectrum.
A similar knee was first expected at $37^{+15}_{-13} \; \mathrm{nHz}$ by \citet{sesana_stochastic_2008}, and is also subtly visible, though not highlighted, in simulated strain spectra from \citet{ravi_does_2012}, \citet{roebber_cosmic_2016}, \citet{kelley_gravitational_2017}, and \citet{taylor_constraints_2017}.
The fact that the double power-law model and the physical model predict different values for $f_{\mathrm{knee}}$ suggest this is somewhat model dependent. However the uncertainties in $f_{\mathrm{knee}}$ for both models are broad, and the median value of $f_{\mathrm{knee}}$ predicted with the physical model lies within the $68\%$ confidence interval of the $f_{\mathrm{knee}}$ predicted with the double power-law model.
It would be interesting to explore systematic differences between these models in future work to aid interpretation of $f_{\mathrm{knee}}$.
The $8^{+8}_{-4} \; \mathrm{nHz}$ knee predicted by the physically motivated model suggests that a knee in the GWB characteristic strain spectrum could be by characterized by future PTA datasets.

We next assessed the level of significance of excursions from an average $f^{-2 / 3}_\mathrm{GW}$ power-law in NANOGrav's 15 yr GWB residual spectrum \citep{agazie_nanograv_2023}.
To determine the significance of these excursions we compared them to the distribution of 1,000 realizations of GWB residual spectra.
We paid particular attention to the frequencies $2 \; \mathrm{nHz}$, $12 \; \mathrm{nHz}$, $14 \; \mathrm{nHz}$, and $16 \; \mathrm{nHz}$, which were previously noted as potentially interesting in \citealt{agazie_nanograv_2023}.
We found that the excursion from an average expected $f_{\mathrm{GW}}^{-2 / 3}$ power-law seen at $2 \; \mathrm{nHz}$ is below the median value of our GWB realizations with $p = 0.05$, the excursions at $12 \; \mathrm{nHz}$ and $14 \; \mathrm{nHz}$ are below the median with $p = 0.31$, and the power at $16 \; \mathrm{nHz}$ is above the median with $p = 0.15$ (\autoref{tab:significance}).

We also find, however, that constraints on the residual spectrum do not strongly rule out weak GW signals at most frequencies.
Consequentially, the significance of excursions from an $f_{\mathrm{GW}}^{-2 / 3}$ power-law depends on the priors.
By cutting the residual spectrum PDFs at higher residuals we find the excursion at $2 \; \mathrm{nHz}$ below the median value of our GWB realizations with $p = 0.05$ to $0.06$, the excursions at $12 \; \mathrm{nHz}$ and $14 \; \mathrm{nHz}$ are consistent with the median value of our GWB realizations with $p = 0.78$ to $0.92$, and the excursion at $16 \; \mathrm{nHz}$ is above the median of our GWB realizations with $p = 0.04$ to $0.05$.

We additionally compared the spectra measured by NANOGrav in the 12.5 yr and 15 yr datasets, re-computing the 15 yr spectrum on the coarser 12.5 yr frequency intervals for a more consistent comparison.
Both data sets show excursions from an average expected $f_{\mathrm{GW}}^{-2 / 3}$ power-law at $\sim 16 \; \mathrm{nHz}$.
The constraints on the GWB spectrum at this frequency are very consistent from the 12.5 yr data set to the 15 yr data set and lie $\sim 2 \sigma$ above the median value of our GWB realizations.
Below $\sim 16 \; \mathrm{nHz}$ there are noticeable differences between both sets of spectra, with the 15 yr spectrum appearing to be better constrained than the 12.5 yr spectrum at these frequencies.
Above $16 \; \mathrm{nHz}$ both spectra appear fully consistent with white noise.

The excursion in the 15 yr GWB spectrum from an $f_{\mathrm{GW}}^{-2 / 3}$ power-law at $2 \; \mathrm{nHz}$ is below our GWB realizations with $p = 0.05$ to $0.06$ ($\approx 1.8 \sigma - 1.9 \sigma$).
This may indicate a turnover in the GWB spectrum due to environmental coupling between SMBHBs and their host environments at low-frequencies, see e.g. \cite{sampson_constraining_2015}.
This scenario has previously been shown to be consistent with the measured GWB spectrum \citep{agazie_nanograv_2023,agazie_nanograv_2023f}.
Specifically, \citet{agazie_nanograv_2023f} report that the $2 \; \mathrm{nHz}$ excursion is consistent with a turnover in the characteristic strain spectrum due to interactions between SMBHBs and their host galaxy environments.
We thus focus our discussion on the $16 \; \mathrm{nHz}$ excursion.

One plausible explanation for the excursion from an expected average $f_{\mathrm{GW}}^{-2 / 3}$ power-law at $16 \; \mathrm{nHz}$ is the existence of a single sufficiently massive and/or nearby SMBHB emitting at $\sim 16 \; \mathrm{nHz}$.
This interpretation is consistent with a GWB sourced by a discrete population of SMBHBs, as an $f^{-2 / 3}_\mathrm{GW}$ power-law only holds on average over many realizations of the GWB.
Indeed, we find that any random realization of the GWB can include large excursions from an $f_{\mathrm{GW}}^{-2 / 3}$ power-law if the underlying SMBHB population includes even one more SMBHB at a higher mass than the analytic average expectation (\autoref{fig:example_spectrum}).

Assuming the excursion at $16 \; \mathrm{nHz}$ is due to a single loud binary, we decomposed the excess strain (compared to the distribution of generated GWB spectra) at this frequency into the corresponding $\mathcal{M}$ and $D_{L}$ (\autoref{fig:bin_8}).
This is consistent with a GWB sourced by a population of discrete SMBHBs, as individual realizations of the GWB do not need to adhere to an average $f^{-2 / 3}_\mathrm{GW}$ power-law.
Interestingly, if the excursion from an $f_{\mathrm{GW}}^{-2 / 3}$ power-law at $16 \; \mathrm{nHz}$ is a single binary it would not be detected in NANOGrav's 15 yr CW search.
The $16 \; \mathrm{nHz}$ excursion could alternatively be sourced by multiple binaries, which would also be undetectable as CWs in NANOGrav's 15 yr data.
This scenario may be less likely than a single loud SMBHB, however, as it would require multiple very massive SMBHBs -- which are expected to be rare (\autoref{fig:discretized_model}) -- to be coincidentally emitting at similar frequencies.
For example, if we assume that HS 1630+2355, HS 0926+3608, and SDSS J114857.33+1600 are all SMBHBs emitting near $16 \; \mathrm{nHz}$, the characteristic strain of their combined GW emission would still only be $\left(6.6^{+8.3}_{-3.6}\right) \times 10^{-16}$ -- which could not source the $16 \; \mathrm{nHz}$ excursion.

One promising avenue for followup are targeted CW searches, which can be up to an order of magnitude more sensitive than all-sky CW searches \citep{arzoumanian_multimessenger_2020}.
Targeted searches for CWs in the 15 yr data set are currently underway.
These searches target SMBHB candidates which have been identified electromagnetically via, e.g., apparent quasar light-curve periodicity \citep{dorazio_accretion_2013,farris_binary_2014,miranda_viscous_2017,munoz_circumbinary_2020}.
Existing SMBHB candidate catalogs may contain SMBHBs with orbital periods $P_{\mathrm{orb}} \lesssim 6 \; \mathrm{yrs}$ \citep{graham_systematic_2015}, corresponding to $f_{\mathrm{GW}} \gtrsim 10 \; \mathrm{nHz}$ (since $f_{\mathrm{GW}} = 2 / P_{\mathrm{orb}}$).
Thus a SMBHB with $f_{\mathrm{GW}} \sim 16 \; \mathrm{nHz}$ may already be present in the extant SMBHB candidate catalogs.

Additionally, the signal-to-noise ratio of the GWB improves proportionally to the number of pulsars in the array \citep{siemens_stochastic_2013}.
Constraints on the GWB spectrum at $16 \; \mathrm{nHz}$ -- which corresponds to GW periods of $\sim 2 \; \mathrm{yrs}$ --  can thus be improved by adding pulsars with $\gtrsim 2 \; \mathrm{yrs}$ of timing data to future NANOGrav data sets.
The International PTA (IPTA), which combines data from PTA experiments around the globe \citep{verbiest_international_2016,perera_international_2019}, could accomplish this in its anticipated third data release by adding data from the MeerKAT PTA, which included $78$ pulsars with $\sim 2.5\; \mathrm{yrs}$ of pulsar timing data in its first data release \citep{miles_meerkat_2023}.

Finally, we must also consider the possibility that the $16 \; \mathrm{nHz}$ excursion does not have a GW origin at all.
Indeed, neither the PPTA nor the EPTA+InPTA spectra have clear excursions at $16 \; \mathrm{nHz}$ compared to NANOGrav's 15 yr GWB spectrum \citep{reardon_search_2023,eptacollaboration_second_2023,agazie_comparing_2024}.
We thus also considered a phenomenological model of the characteristic strain spectrum that includes $\sim 7 \; \mathrm{ns}$ of excess white noise in the residual spectrum of the GWB.
We find that the excursion at $16 \; \mathrm{nHz}$ is above our generated GWB realizations with a reduced significance of $p = 0.3 - 0.4$, corresponding to a $\sim 1 \sigma$ excursion.
However we emphasize that current NANOGrav analyses already consider white noise when fitting the pulsar timing model to timing residual data, with a reduced $\chi^{2}$ near unity.

While searching for CWs in the IPTA's second data release, \citet{falxa_searching_2023} found that a one-size-fits-all approach to pulsar noise modelling can increase the probability of a false alarm.
They further found that custom pulsar noise models reduce false alarms and improve our ability to constrain CWs from individual SMBHBs. 
PTA experiments could test whether or not the $16 \; \mathrm{nHz}$ excursion is due to mis-modeled pulsar noise by incorporating custom pulsar noise models in future data sets \citep{lentati_spin_2016,chalumeau_noise_2022,larsen_nanograv_2024,goncharov_fewer_2024}.
It will be interesting to see how custom noise models for individual pulsars in the 15 yr data set will affect apparent white noise levels (Agazie et al. in prep.), and thus potentially change our results.
Determining the specific source of the $16 \; \mathrm{nHz}$ power-law excursion -- including whether or not it is due to a GW source at all -- will require further investigation.

Our study is also important to help assess if SMBHBs may indeed be likely sources of the GWB, alongside searches for CWs \citep{arzoumanian_nanograv_2023a,agazie_nanograv_2023d} and anisotropy \citep{mingarelli_characterizing_2013,taylor_searching_2013,cornish_mapping_2014,gair_mapping_2014,taylor_limits_2015,taylor_bright_2020,pol_forecasting_2022,agazie_nanograv_2023c,gardiner_background_2024,sato-polito_exploring_2024}.
Without a careful exploration of the strain spectrum it will not be clear if excursions from the expected average power-law behavior are consistent with what we expect from a population of discrete SMBHBs, or are sourced by some other physics, e.g. \citet{afzal_nanograv_2023}.
Indeed, primordial GWB are expected to be isotropic and also follow a power-law \citep{grishchuk_reviews_2005,lasky_gravitationalwave_2016,afzal_nanograv_2023,agazie_nanograv_2024d}.
Signs of discreteness in the GWB residual spectrum are therefore an important signature of SMBHBs, and may be observed before GWB anisotropy or CWs.

We show conclusively that excursions in NANOGrav's 15 yr GWB spectrum are well within the range of spectra  generated by a population of discrete SMBHBs, needing no other physics to explain it.
Furthermore, we predict that the GWB breaks down at $26^{+28}_{-19} \; \mathrm{nHz}$.
If white noise can be reduced, the rest of the available high-frequency parameter space will be ideal for constraining CWs.
Finally our study also highlights the importance of spectral analyses such as those carried out by NANOGrav, which reveal interesting features in the GWB.
Recent advancements in spectral analysis techniques have reduced the computational cost associated with constraining the GWB spectrum, making spectral analyses more accessible for PTAs to carry out, e.g. \cite{lamb_rapid_2023}.

\emph{Author Contributions:}
This paper is the result of the work of many people and uses data from over a decade of pulsar timing observations.
J.A.C.C. wrote and developed new python codes to perform the analysis, created all the figures and tables, and wrote a majority of the text.
C.M.F.M. conceived of the project, supervised the analysis, helped write and develop the manuscript, and advised on the analysis, the figures, and interpretation of the results. W.G.L. resampled the 15 yr data in Figure 6, left hand side.
L.Z.K, K.D.O, L.Br., P.N., B.B., S.V., and D.K. provided insights into interpreting the results, and comments on the manuscript. 

G.A., A.A, A.M.A., Z.A., P.T.B., P.R.B., H.T.C., K.C., M.E.D, P.B.D., T.D., E.C.F, W.F., E.F., G.E.F., N.G.D., D.C.G., P.A.G., J.G., R.J.J., M.L.J., D.L.K., M.K., M.T.L., D.R.L., J.L., R.S.L., A.M., M.A.M., N.M., B.W.M., C.N., D.J.N., T.T.N., B.B.P.P., N.S.P., H.A.R., S.M.R., P.S.R., A.S., C.S., B.J.S.A., I.H.S., K.S., A.S., J.K.S., and H.M.W. all ran observations and developed timing models for the NANOGrav 15 yr data set.

\textit{Acknowledgements:} L.Bl.\ acknowledges support from the National Science Foundation under award AST-1909933 and from the Research Corporation for Science Advancement under Cottrell Scholar Award No.\ 27553.
P.R.B.\ is supported by the Science and Technology Facilities Council, grant number ST/W000946/1.
S.B.\ gratefully acknowledges the support of a Sloan Fellowship, and the support of NSF under award \#1815664.
M.C.\ and S.R.T.\ acknowledge support from NSF AST-2007993.
M.C.\ was supported by the Vanderbilt Initiative in Data Intensive Astrophysics (VIDA) Fellowship.
Support for this work was provided by the NSF through the Grote Reber Fellowship Program administered by Associated Universities, Inc./National Radio Astronomy Observatory.
M.E.D.\ acknowledges support from the Naval Research Laboratory by NASA under contract S-15633Y.
T.D.\ and M.T.L.\ are supported by an NSF Astronomy and Astrophysics Grant (AAG) award number 2009468.
E.C.F.\ is supported by NASA under award number 80GSFC21M0002.
G.E.F., S.C.S., and S.J.V.\ are supported by NSF award PHY-2011772.
K.A.G.\ and S.R.T.\ acknowledge support from an NSF CAREER award \#2146016.
A.D.J.\ and M.V.\ acknowledge support from the Caltech and Jet Propulsion Laboratory President's and Director's Research and Development Fund.
A.D.J.\ acknowledges support from the Sloan Foundation.
The work of N.La., X.S., and D.W.\ is partly supported by the George and Hannah Bolinger Memorial Fund in the College of Science at Oregon State University.
N.La.\ acknowledges the support from Larry W. Martin and Joyce B. O'Neill Endowed Fellowship in the College of Science at Oregon State University.
Part of this research was carried out at the Jet Propulsion Laboratory, California Institute of Technology, under a contract with the National Aeronautics and Space Administration (80NM0018D0004).
M.A.M.\ is supported by NSF \#1458952.
M.A.M.\ is supported by NSF \#2009425.
C.M.F.M.\ was supported in part by the National Science Foundation under Grants No.\ NSF PHY-1748958 and AST-2106552.
A.Mi.\ is supported by the Deutsche Forschungsgemeinschaft under Germany's Excellence Strategy - EXC 2121 Quantum Universe - 390833306.
K.D.O.\ was supported in part by NSF Grant No.\ 2207267.
T.T.P.\ acknowledges support from the Extragalactic Astrophysics Research Group at E\"{o}tv\"{o}s Lor\'{a}nd University, funded by the E\"{o}tv\"{o}s Lor\'{a}nd Research Network (ELKH), which was used during the development of this research.
H.A.R.\ is supported by NSF Partnerships for Research and Education in Physics (PREP) award No.\ 2216793.
S.M.R.\ and I.H.S.\ are CIFAR Fellows.
Portions of this work performed at NRL were supported by ONR 6.1 basic research funding.
J.D.R.\ also acknowledges support from start-up funds from Texas Tech University.
J.S.\ is supported by an NSF Astronomy and Astrophysics Postdoctoral Fellowship under award AST-2202388, and acknowledges previous support by the NSF under award 1847938.
Pulsar research at UBC is supported by an NSERC Discovery Grant and by CIFAR.
C.U.\ acknowledges support from BGU (Kreitman fellowship), and the Council for Higher Education and Israel Academy of Sciences and Humanities (Excellence fellowship).
C.A.W.\ acknowledges support from CIERA, the Adler Planetarium, and the Brinson Foundation through a CIERA-Adler postdoctoral fellowship.
O.Y.\ is supported by the National Science Foundation Graduate Research Fellowship under Grant No.\ DGE-2139292.

\software{
          arviz \citep{kumar_arviz_2019},
          astropy \citep{astropycollaboration_astropy_2013,astropycollaboration_astropy_2018,astropycollaboration_astropy_2022},
          jupyter \citep{kluyver_jupyter_2016},
          matplotlib \citep{hunter_matplotlib_2007},
          nHzGWs \citep{mingarelli_chiaramingarelli_2017},
          numpy \citep{harris_array_2020},
          pandas \citep{mckinney_data_2010,thepandasdevelopmentteam_pandasdev_2022},
          pymc \citep{wiecki_pymcdevs_2023},
          scikit-learn \citep{pedregosa_scikitlearn_2011},
          scipy \citep{virtanen_scipy_2020},
          seaborn \citep{waskom_seaborn_2021}
          }
          
\clearpage
\appendix

\section{Broken Power-Law}
\label{app:broken_pl}

\begin{figure}[t!]
    \centering
    \includegraphics{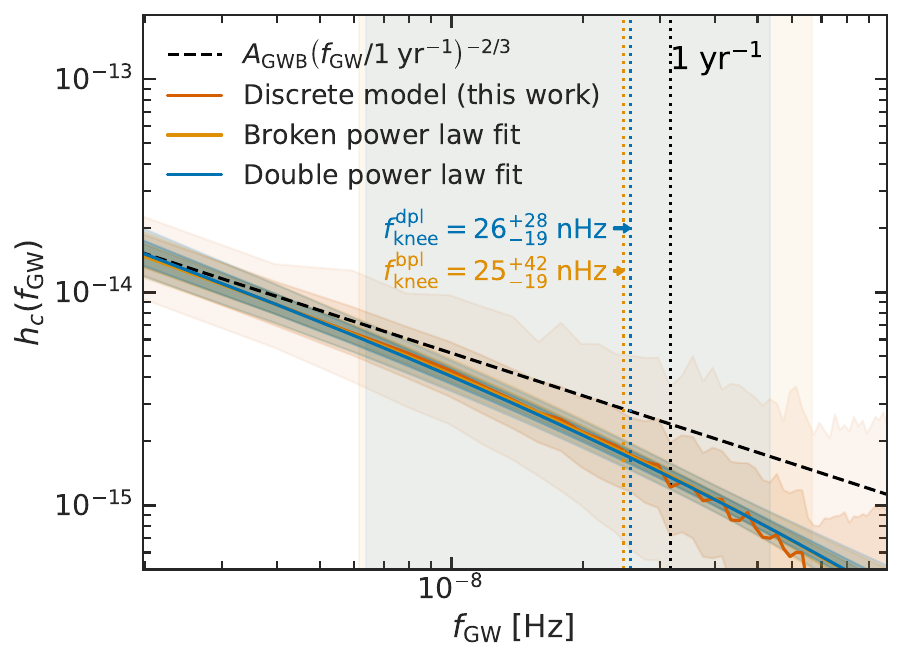}
    \caption{The broken power-law model (yellow) is nearly identical to the double power-law model (blue).}
    \label{fig:both_power_laws}
\end{figure}

Here we fit a broken power-law model to our simulated GWB spectra.
This model was first employed by \citet{sesana_stochastic_2008} to fit the knee of GWB realizations derived from halo merger trees.
The broken power-law characteristic strain spectrum model is:
\begin{equation}
\label{eq:bpl}
    h_{c}(f_{\mathrm{GW}}) = A_{\rm yr} \left(\frac{f_{\mathrm{GW}}}{f_{\rm yr}}\right)^{\alpha_{1}} \left(1 + \frac{f_{\mathrm{GW}}}{f_{\rm knee}}\right)^{\alpha_{2} - \alpha_{1}}.
\end{equation}
Here $\alpha_{1}$ and $\alpha_{2}$ are the low- and high-frequency characteristic strain spectrum slopes, respectively. $\alpha_{1} = -2 / 3$ for circular SMBHBs. Fitting this to our GWB realizations we find a bend at $f_{\mathrm{knee}} = 25^{+42}_{-19} \; \mathrm{nHz}$ with a high frequency power-law slope $f_{\mathrm{GW}}^{-1.4^{+0.3}_{-0.5}}$. The results of this fit are shown in yellow in \autoref{fig:both_power_laws}, and are nearly identical to the double power-law fit.

\section{Physical Spectrum}
\label{sec:physical_model}

Here we consider a physically motivated model of $h_{c}(f_{\mathrm{GW}})$ which accounts for the fact that the total number of binaries in any frequency interval must be an integer, i.e., one cannot have fractions of a SMBHB contributing to the GWB \citep{sesana_stochastic_2008}.
Specifically, the total number of SMBHBs in frequency interval $\Delta f_{\mathrm{GW}}$ centered on frequency $f_{0}$ is
\begin{equation}
\label{eq:N_f}
    N_{0} = \int_{0}^{\infty} \int_{f_{0} - \Delta f_{\mathrm{GW}} / 2}^{f_{0} + \Delta f_{\mathrm{GW}} / 2} \frac{\dd[2]{N_{\mathrm{BHB}}}}{\dd{M} \dd{f_{\mathrm{GW}}}} \dd{M} \dd{f_{\mathrm{GW}}} \, ,
\end{equation} 
where $\dd[2]{N_{\mathrm{BHB}}} / (\dd{M} \dd{f_{\mathrm{GW}}}) = \int_{0}^{\infty} \int_{0}^{1} \dd[2]{N_{\mathrm{BHB}}} / (\dd{M} \dd{f_{\mathrm{GW}}}) \dd{q} \dd{z}$.
In general, $N_{0}$ is not an integer. This is because $N_{0}$ is really the expected number of binaries in $\Delta f_{\mathrm{GW}}$ over many realizations of the universe.

\citet{sesana_stochastic_2008} therefore posit a characteristic mass, $\tilde{M}(f_{0})$, such that over infinite realizations of the universe there is less than one SMBHB with $M > \tilde{M}(f_{0})$ on average, i.e., 
\begin{equation}
\label{eq:char_mass}
\begin{split}
    \int_{\tilde{M}(f_{0})}^{\infty} \int_{f_{0} - \Delta f_{\mathrm{GW}} / 2}^{f_{0} + \Delta f_{\mathrm{GW}} / 2} \frac{\dd[2]{N_{\mathrm{BHB}}}}{\dd{M} \dd{f_{\mathrm{GW}}}} \dd{M} \dd{f_{\mathrm{GW}}} < 1 \\
    \Rightarrow \int_{\tilde{M}(f_{0})}^{\infty} p(M) \dd{M} \int_{f_{0} - \Delta f_{\mathrm{GW}} / 2}^{f_{0} + \Delta f_{\mathrm{GW}} / 2} \frac{\dd{N_{\mathrm{BHB}}}}{\dd{f_{\mathrm{GW}}}} \dd{f_{\mathrm{GW}}} < 1
\end{split} \, ,
\end{equation} 
where in the second line $\dd{N_{\mathrm{BHB}}} / \dd{f_{\mathrm{GW}}}$ is the differential number of SMBHBs per $f_{\mathrm{GW}}$ and $p(M)$ is the probability density function of SMBHBs with mass $M$, such that $\dd[2]{N_{\mathrm{BHB}}} / (\dd{M} \dd{f_{\mathrm{GW}}}) = p(M) \times \dd{N_{\mathrm{BHB}}} / \dd{f_{\mathrm{GW}}}$.
SMBHBs with $M > \tilde{M}(f_{0})$ thus do not contribute to the GWB in a typical realization of the universe.
To account for this discreteness effect we model the effective characteristic strain spectrum as
$h_{c, \mathrm{eff}}(f_{\mathrm{GW}}) = h_{c}(f_{\mathrm{GW}}) \sqrt{1 - Z(f_{\mathrm{GW}})}$, where
\begin{equation}
\label{eq:missing_frac}
    Z(f_{\mathrm{GW}}) = \frac{\int_{\tilde{M}(f_{\mathrm{GW}})}^{\infty} \int_{0}^{1} \int_{0}^{\infty} \dot{\phi}_{\rm BHB} \frac{\dd{t_{r}}}{\dd{z} }\frac{\mathcal{M}^{5 / 3}}{(1 + z)^{1 /3}} \dd{M} \dd{q} \dd{z}}{\int_{0}^{\infty} \int_{0}^{1} \int_{0}^{\infty} \dot{\phi}_{\rm BHB} \frac{\dd{t_{r}}}{\dd{z} }\frac{\mathcal{M}^{5 / 3}}{(1 + z)^{1 /3}} \dd{M} \dd{q} \dd{z}}
\end{equation}
is the fraction of contributions to $h_{c}^{2}(f_{\mathrm{GW}})$ coming from less than one SMBHB \citep{sesana_stochastic_2008}.

We can derive an analytic expression for $Z(f_{\mathrm{GW}})$ -- and thus for $h_{c, \mathrm{eff}}(f_{\mathrm{GW}})$ -- by assuming that $\dot{\phi}_{\mathrm{BHB}}$ is power-law distributed over mass \citep{sesana_stochastic_2008}.
Indeed, halo merger tree models of SMBHBs \citep[e.g.][]{volonteri_assembly_2003,sesana_stochastic_2008} find $\dd[2]{N}_{\mathrm{BHB}} / (\dd{M} \dd{t_{\mathrm{r}}}) \propto M^{-\beta}$, with $1.5 \lesssim \beta \lesssim 2$ \citep{sesana_stochastic_2008}.
We also find a similar distribution in our generated SMBHB populations.
Since $\dot{\phi}_{\mathrm{BHB}} \propto \dd[2]{N}_{\mathrm{BHB}} / (\dd{M} \dd{t_{\mathrm{r}}})$ (\autoref{eq:diff_num}), this suggests that $\dot{\phi}_{\mathrm{BHB}}$ can also be approximated as a power-law, i.e.,
\begin{equation}
\label{eq:smbhb_merger_rate_approx}
    \dot{\phi} \approx \begin{cases}
        A(q, z) M^{-\beta} & M_{\min} \leq M \leq M_{\max} \\
        0 & \text{otherwise}
    \end{cases} \, ,
\end{equation}
where $A(q, z)$ is a model dependent normalization factor which does not depend on $M$ (and therefore cancels out in \autoref{eq:missing_frac}), and where $M_{\min} = 10^{8.5} \; \mathrm{M}_{\odot}$ and $M_{\max} = 10^{10.5} \; \mathrm{M}_{\odot}$ are the bounds of integration. 
Plugging \autoref{eq:smbhb_merger_rate_approx} into \autoref{eq:missing_frac}, we find \citep[cf.][]{sesana_stochastic_2008}
\begin{equation}
\label{eq:z_approx}
\begin{split}
    Z(f_{\mathrm{GW}}) &= \frac{\int_{\tilde{M}(f_{\mathrm{GW}})}^{M_{\max}} M^{5 / 3 - \beta} \dd{M}}{\int_{M_{\min}}^{M_{\max}} M^{5 / 3 - \beta} \dd{M}} \\
    &\simeq 1 - \left(\frac{\tilde{M}(f_{\mathrm{GW}})}{M_{\max}}\right)^{8/3 - \beta} \, ,
\end{split}
\end{equation}
where in the second line of \autoref{eq:z_approx} we have assumed $\beta < 8/3$ and $\tilde{M}(f_{\mathrm{GW}}) \gg M_{\min}$.
This is true for our generated populations at most frequencies and is in general true for a real population of SMBHBs, which can include small contributions to the characteristic strain from lower mass SMBHBs than those considered in this work \citep{casey-clyde_quasarbased_2022}.

We next solve for $\tilde{M}(f_{\mathrm{GW}})$ before finally arriving at an analytic expression for $Z(f_{\mathrm{GW}})$.
Since $\dd{t_{\mathrm{r}}} / \dd{f_{\mathrm{GW}}} \propto M^{-5 / 3} f_{\mathrm{GW}}^{-11 / 3}$ \citep{peters_gravitational_1963}, it follows from our power-law formulation of $\dot{\phi}_{\mathrm{BHB}}$ that $\dd[2] N_{\mathrm{BHB}} / (\dd{M} \dd{f_{\mathrm{GW}}}) \propto M^{-\beta -5 / 3} f_{\mathrm{GW}}^{-11 / 3}$ \citep{sesana_stochastic_2008}.
The total number of SMBHBs per frequency interval $\Delta f_{\mathrm{GW}}$ centered on frequency $f_{\mathrm{GW}}$ is thus $N(f_{\mathrm{GW}}) = N_{0} \left(f_{\mathrm{GW}} / f_{0}\right)^{-11/3}$ \citep{sesana_stochastic_2008} where $f_{0}$ is an arbitrary reference frequency and $N_{0}$ is set by \autoref{eq:N_f}.
Combining this expression for $N(f_{\mathrm{GW}})$ with \cref{eq:N_f,eq:char_mass} and solving for $\tilde{M}$ we find \citep[cf.][]{sesana_stochastic_2008}
\begin{equation}
\label{eq:char_mass_solved}
    \tilde{M}(f_{\mathrm{GW}}) = M_{\max} \left[1 + \frac{\beta + 2/3}{B N_{0} M_{\max}^{-\beta - 2/3}} \left(\frac{f_{\mathrm{GW}}}{f_{0}}\right)^{11 / 3}\right]^{1 / (-\beta - 2/3)} \, ,
\end{equation}
where $B$ is a normalization constant such that $\int_{0}^{\infty} p(M) \dd{M} = \int_{0}^{\infty} B M^{-\beta - 5 / 3} \dd{M} = 1$.
Combining \cref{eq:z_approx,eq:char_mass_solved}, and defining $f_{\mathrm{knee}} \equiv f_{0} \left[(\beta + 2/3)/(B N_{0} M_{\max}^{-\beta - 2/3})\right]^{-3/11}$, the characteristic strain spectrum can be parameterized as
\begin{equation}
\label{eq:phys_app}
    h_{c, \mathrm{eff}} = A_{\mathrm{yr}} \left(\frac{f_{\mathrm{GW}}}{f_{\mathrm{yr}}}\right)^{\alpha_{1}} \left[1 + \left(\frac{f_{\mathrm{GW}}}{f_{\mathrm{knee}}}\right)^{11/3}\right]^{3 (\alpha_{2} - \alpha_{1}) / 11} \, ,
\end{equation}
where $\alpha_{1} = -2/3$ is the slope at $f_{\mathrm{GW}} \ll f_{\mathrm{knee}}$, and where $\alpha_{2} = \alpha_{1} + (11 / 6) (3 \beta - 8) / (3 \beta + 2)$ is the slope of the average strain spectrum at $f_{\mathrm{GW}} \gg f_{\mathrm{knee}}$.
At $f_{\mathrm{GW}} \gg f_{\mathrm{knee}}$, we therefore have $h_{c, \mathrm{eff}} \propto f_{\mathrm{GW}}^{-2/3 + (11 / 6) (3 \beta - 8) / (3 \beta + 2)}$, which for $1.5 \lesssim \beta \lesssim 2$ is $h_{c, \mathrm{eff}} \propto f_{\mathrm{GW}}^{-1.6} - f_{\mathrm{GW}}^{-1.1}$, in good agreement with both the double and broken power-law models.
In fact by fitting \autoref{eq:phys_app} to the 1,000 spectra we generate from discrete SMBHB populations we find $h_{c} \propto f_{\mathrm{GW}}^{-1.1^{+0.1}_{-0.2}}$. This corresponds to $\beta = 2.1^{+0.1}_{-0.2}$, which is consistent with expectations from halo merger tree models \citep{sesana_stochastic_2008} and the SMBHB populations used in this work.

\bibliography{bib}

\end{document}